\documentclass[]{aa}
\pdfoutput=1
\usepackage{txfonts}
\usepackage{natbib}
\usepackage{graphicx}
\usepackage{longtable}
\bibpunct{(}{)}{;}{a}{}{,}
\begin{document}

\title{Three Galactic globular cluster candidates\thanks{Based
on observations gathered with ESO-VISTA telescope (proposal ID 172.B-2002).}}
\subtitle{}

\author{
C. Moni Bidin \inst{1}
\and
F. Mauro \inst{1}
\and
D. Geisler \inst{1}
\and
D. Minniti \inst{2}
\and
M. Catelan \inst{2}
\and
M. Hempel \inst{2}
\and
E. Valenti \inst{3}
\and
A.~A.~R. Valcarce \inst{2,5}
\and
J. Alonso-Garc\'ia \inst{2}
\and
J. Borissova \inst{6}
\and
G. Carraro \inst{4}
\and
P. Lucas \inst{7}
\and
A.-N. Chen\'e \inst{1,6}
\and
M. Zoccali \inst{2}
\and
R. G. Kurtev \inst{6}
}

\institute{
Departamento de Astronom\'ia, Universidad de Concepci\'on, Casilla 160-C, Concepci\'on, Chile
\and
Departamento de Astronom\'ia y Astrof\'isica, Pontificia Universidad Cat\'olica de Chile,
Casilla 306, Santiago, Chile
\and
European Southern Observatory, Karl\--Schwarzschild\--Stra\ss e 2, D\--85748 Garching bei M\"{u}nchen, Germany.
\and
European Southern Observatory, Avda Alonso de Cordova, 3107, Casilla 19001, Santiago, Chile
\and
Departamento de F\'isica, Universidade Federal do Rio Grande do Norte, 59072-970 Natal, RN, Brazil
\and
Departamento de F\'isica y Astronom\'ia, Facultad de Ciencias, Universidad de Valpara\'iso,
Av. Gran Breta\~na 1111, Valpara\'iso, Chile
\and
Centre for Astrophysics Research, Science and Technology Research Institute, University of
Hertfordshire, Hatfield AL10 9AB, UK}
\date{Received / Accepted }


\abstract
{The census of Galactic globular clusters (GCs) is still incomplete, and about ten new objects are supposed
to await discovery, hidden behind the crowded and reddened regions of the Galactic bulge and disk.}
{We investigated the nature of three new GC candidates, discovered in the frames collected by the Vista Variables
in the Via Lactea (VVV) near-infrared survey. They will be called VVV~CL002, VVV~CL003, and VVV~CL004.}
{We studied the results of point-spread-function near-infrared photometry from VVV data for the three objects and
their surrounding fields, the proper motion information available in the literature and, when possible, we derived
the cluster parameters by means of calibrated indices measured on the color-magnitude diagrams.}
{The evidence shows that VVV~CL002 is a newly discovered, small, moderately metal-rich
($[\mathrm{Fe/H}]\approx -0.4$) Galactic GC. It is located at a
Galactocentric distance of 0.7$\pm$0.9~kpc, and it could be one of the nearest GC to the Galactic center. Its
characteristics are more similar to those of low-mass, Palomar-like GCs than to more classical, old, and massive
bulge GCs. VVV~CL003 is the first star cluster discovered in the Galactic disk on the opposite side of the center
with respect to the Sun, at a Galactocentric distance of $\sim$5~kpc. Its high metallicity
($[\mathrm{Fe/H}]\approx -0.1$) and location point to an open cluster, but a GC cannot be excluded.
VVV~CL004, on the contrary, is most probably only a random clump of field stars, as indicated by both its low
statistical significance and by the impossibility to distinguish its stars from the surrounding field population.}
{We claim the detection of $i$) a new Galactic GC, deriving an estimate of its basic parameters; $ii$) a
stellar aggregate, probably an open cluster, in the disk directly beyond the Galactic center; and $iii$) 
an overdensity of stars, most probably an asterism.}


\authorrunning{Moni Bidin et al.}
\mail{cmbidin@astro-udec.cl}
\titlerunning{Three Galactic globular cluster candidates}
\maketitle


\section{Introduction}
\label{c_intro}

The census of Galactic globular clusters (GCs) is still incomplete, because distant objects can easily
hide behind the very crowded and highly reddened stellar fields in the direction of the Galactic bulge and
disk. Indeed, \citet{Ivanov05} estimated that there should be 10$\pm$3 undiscovered GCs hiding in the inner
Milky Way. The advent of the new generation of extensive surveys such as SDSS \citep{Abazajian09}, 2MASS
\citep{Skrutskie06}, and GLIMPSE \citep{Benjamin03} permitted the detection of several new Galactic GCs. The
December 2010 compilation of the \citet{Harris96} catalog included seven new GCs not present in the February
2003 version, but seven more objects have been proposed in the last years: SDSSJ1257+3419 \citep{Sakamoto06},
FSR\,584 \citep{Bica07}, FSR\,1767 \citep{Bonatto07}, FSR\,190 \citep{Froebrich08}, Pfleiderer\,2
\citep{Ortolani09}, VVV~CL001 \citep{Minniti11}, and Mercer\,5 \citep{Longmore11}. The family of Galactic GCs
is thus steadily increasing, although the true nature of the GC candidates is not always easy to unveil,
and some objects are still debated \citep[see, for example,][]{Froebrich08b}.

Globular clusters are old objects, whose age ranges from about half to nearly one Hubble time. Their spatial
distribution and physical/chemical properties thus bear information about the Galactic formation and
evolution processes. However, they continuously loose mass during their entire lifetime because of both
external processes such as Galactic tidal stress, and internal ones such as stellar evaporation
\citep{Meylan97,Gnedin97}. The identification and the study of low-mass GCs in the Galactic bulge can
therefore provide new information about both its formation and chemical enrichment history, and the tidal
forces governing its dense environment.

\begin{figure}
\begin{center}
\includegraphics[width=7.5cm]{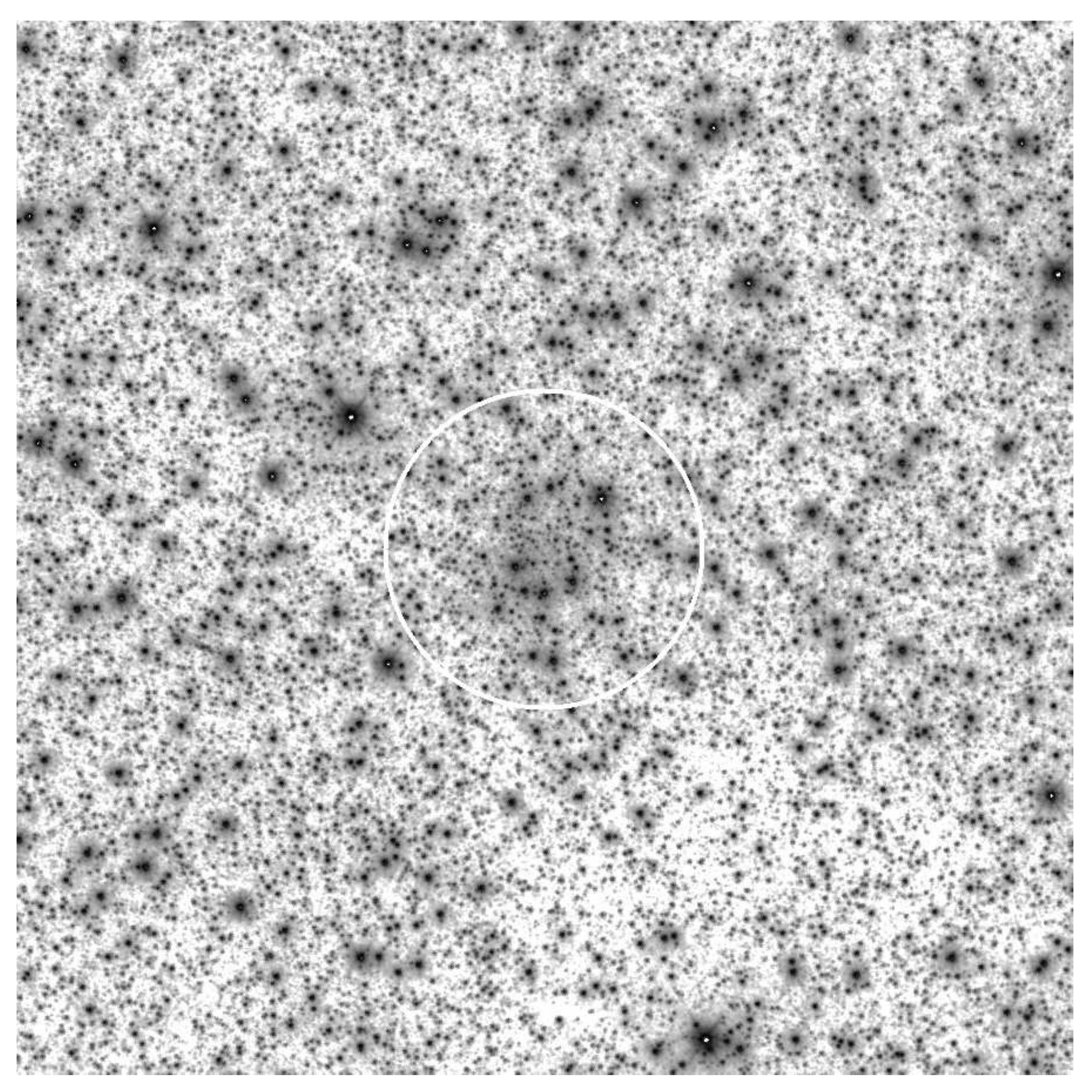}
\includegraphics[width=7.5cm]{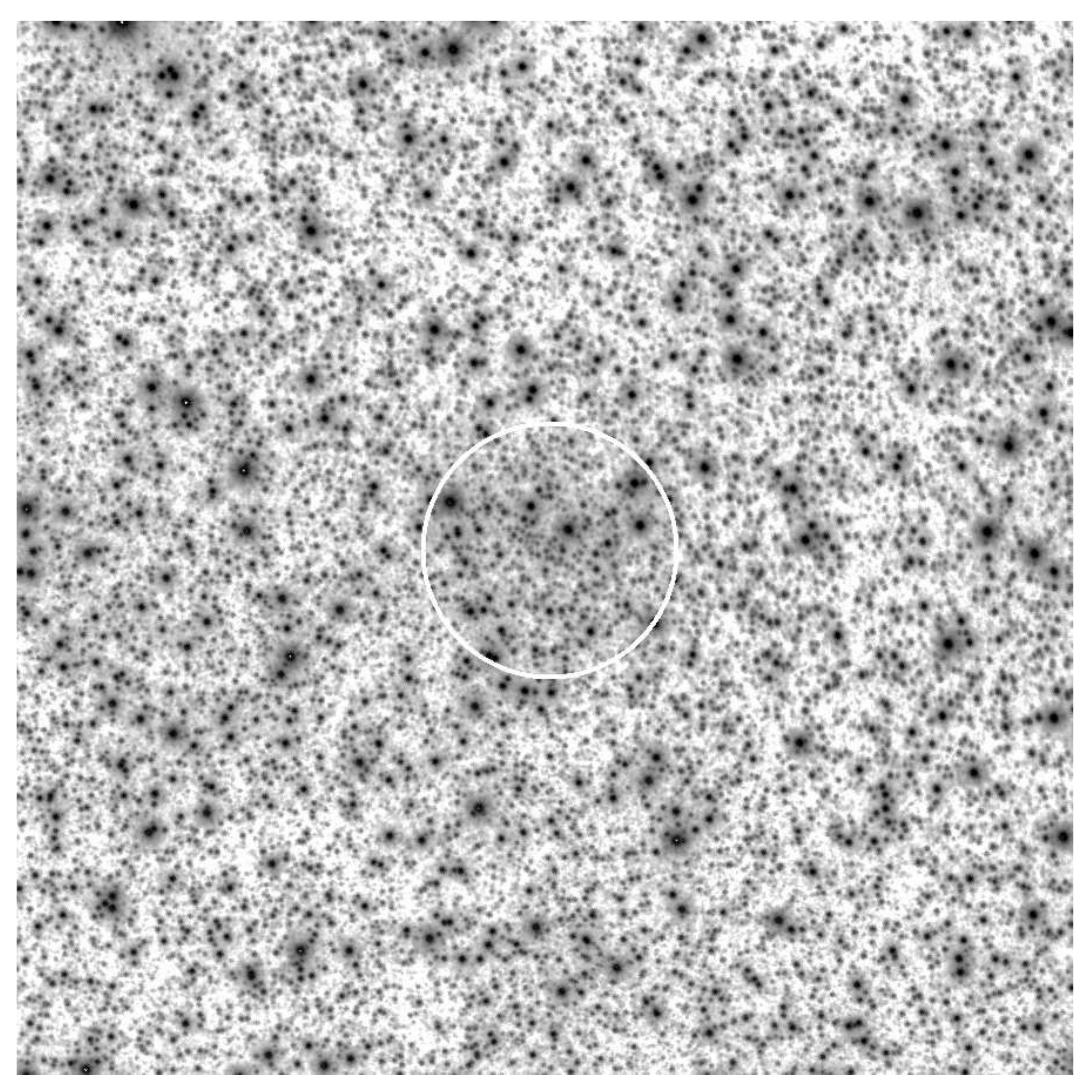}
\includegraphics[width=7.5cm]{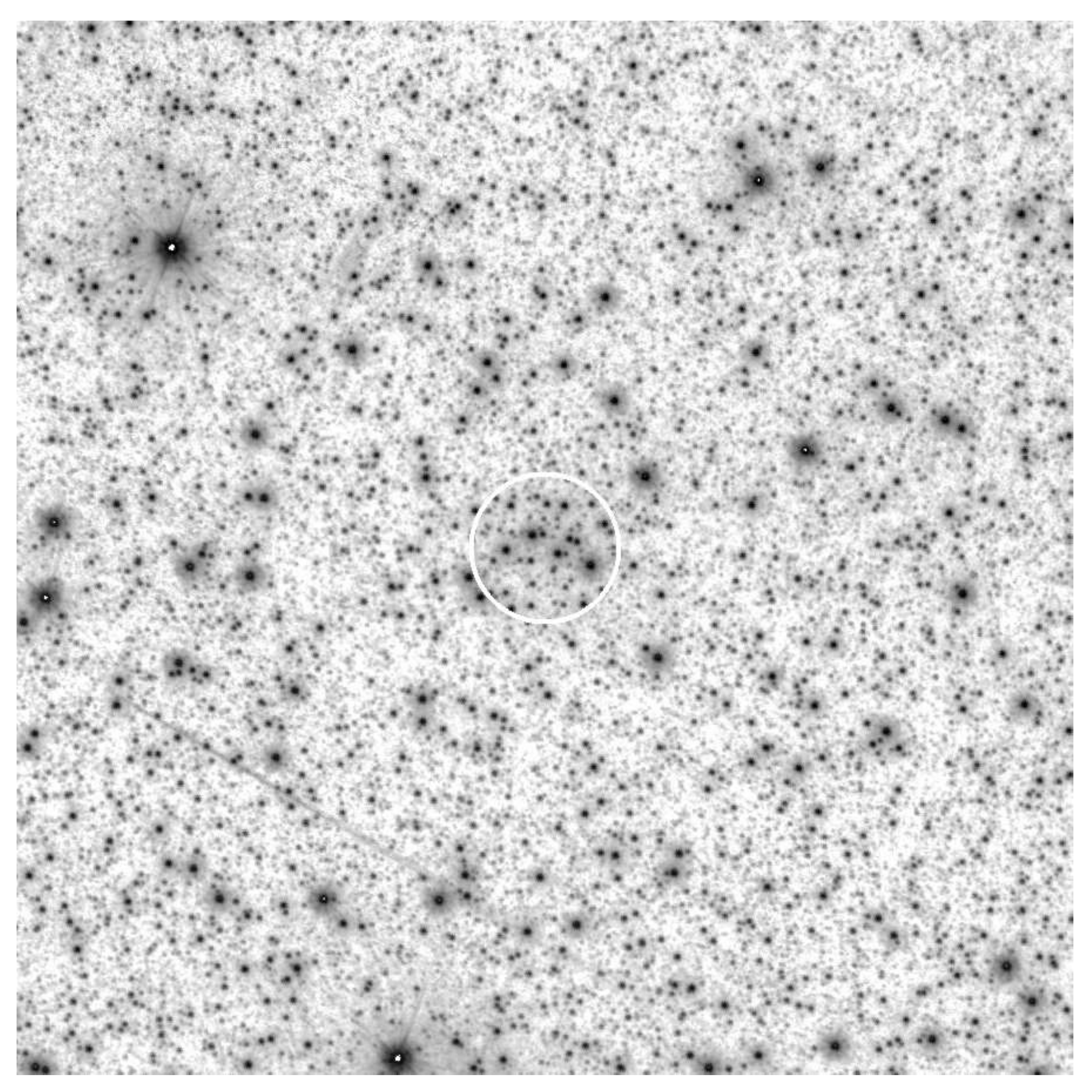}
\caption{5$\arcmin\times 5\arcmin$ logarithmic gray-scale $K_\mathrm{s}$-band images from the VVV archive,
centered on the three candidate clusters (from upper to lower panel: VVV~CL002, VVV~CL003, and VVV~CL004).
North is up and East is to the right. The white circles, centered on the derived cluster center, indicate the
half-light radius estimated for each object.}
\label{f_tile}
\end{center}
\end{figure}

In this paper we analyze three new cluster candidates discovered in the VISTA Variables in the Via Lactea
(VVV) Public Survey, which is gathering near-IR data of the Galactic bulge and inner disk
\citep{Minniti10,Saito10,Catelan11}. This survey is scanning the central regions of the Milky Way with
unprecedented depth and extension, thus generating the ideal database to search for missing GCs in the
inner Galaxy. Following the nomenclature introduced by \citet{Minniti11} and \citet{Borissova11}, the
objects under study here will be called VVV~CL002, VVV~CL003, and VVV~CL004.


\section{Photometric data}
\label{c_data}

\subsection{The cluster candidates}
\label{c_objects}

VVV~CL002, VVV~CL003, and VVV~CL004 were identified as previously unknown stellar overdensities by visual
inspection of the three-band ($J$, $H$, $K_\mathrm{s}$), color composite frames produced by the VVV survey.
Table~\ref{t_coord} reports for each object the corresponding VVV fields and the J2000 coordinates of their
center, as determined in \S\ref{c_cent}. Figure~\ref{f_tile} shows the three objects in the
$K_\mathrm{s}$-band stacked tiles downloaded from the Vista Science Archive
website\footnote{http://horus.roe.ac.uk/vsa/} (VSA). VVV~CL002 and VVV~CL003 look like loose and low-mass,
Palomar-like GCs, although VVV~CL003 is less clear because it is embedded in a very crowded stellar field.
VVV~CL004, on the contrary, does not show a clear concentration of faint stars, but only a local overdensity
of a few bright objects.

\begin{table*}[t]
\begin{center}
\caption{Coordinates of the center of the three GC candidates, and the VVV data used in the present
investigation. The classification (GC= globular cluster, OC= open cluster, NC= not a cluster) resulting
from our investigation is given in the last column.}
\label{t_coord}
\begin{tabular}{c c c c c c c c}
\hline
\hline
ID & RA & dec & $l$ & $b$ & VVV field & frames & classification \\
 & hh:mm:ss & deg:mm:ss & deg & deg & & & \\
\hline
VVV~CL002 & 17:41:06.30 & $-$28:50:42.3 & 359.5586 & 0.8888 & b347 & 2$\times J$, 2$\times H$, 4$\times K_\mathrm{s}$ & GC \\
VVV~CL003 & 17:38:54.56 & $-$29:54:25.3 & 358.4047 & 0.7298 & b346 & 2$\times J$, 2$\times H$, 4$\times K_\mathrm{s}$ & OC? \\
VVV~CL004 & 17:54:31.91 & $-$22:13:37.8 & 6.7900 & 1.7179 & b352  &  4$\times J$, 4$\times H$, 4$\times K_\mathrm{s}$ & NC \\
\hline
\end{tabular}
\end{center}
\end{table*}

\subsection{Data reduction and photometry}
\label{c_reduction}

After the discovery, we retrieved from VSA the stacked images of the individual 2048$\times$2048 pixel
exposures containing the three candidate clusters. The quantity of frames used in each band is indicated in
Table~\ref{t_coord}. The effective exposure time of the $H$- and $K_\mathrm{s}$-band images was 8s, while
it was 24s in $J$. The frames in the $Y$ and $Z$ bands, included in the observation plan of the VVV survey,
were not available at the time of the investigation. The data were acquired on 2010 April 7 to 14, with
the VIRCAM camera mounted on the VISTA 4m telescope at the Paranal Observatory \citep{Emerson10}, and
reduced at the Cambridge Astronomical Survey Unit (CASU)\footnote{http://casu.ast.cam.ac.uk/} with
the VIRCAM pipeline v1.0 \citep{Irwin04}. During the observations the weather conditions fell within the
survey's constraints for seeing, airmass, and Moon distance \citep{Minniti10}, and the quality of the data
was satisfactory.

\begin{figure}
\begin{center}
\resizebox{\hsize}{!}{\includegraphics{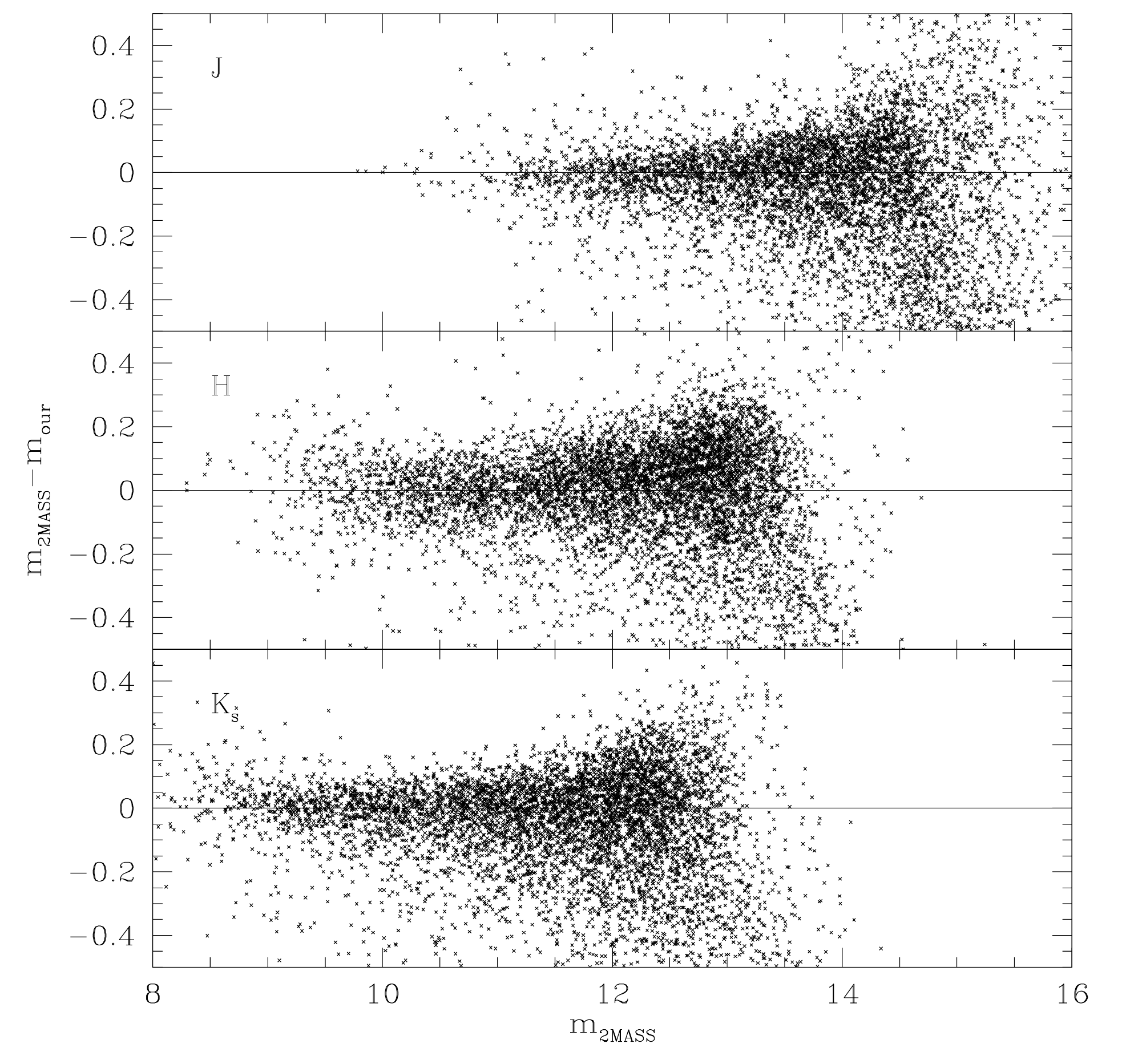}}
\caption{Magnitude difference between 2MASS and our photometry for all the stars in common as a function
of 2MASS magnitude in the stellar field of VVV~CL002. From top to bottom: $J$, $H$, and $K_\mathrm{s}$
magnitudes.}
\label{f_2mas}
\end{center}
\end{figure}

\begin{figure}
\begin{center}
\resizebox{\hsize}{!}{\includegraphics{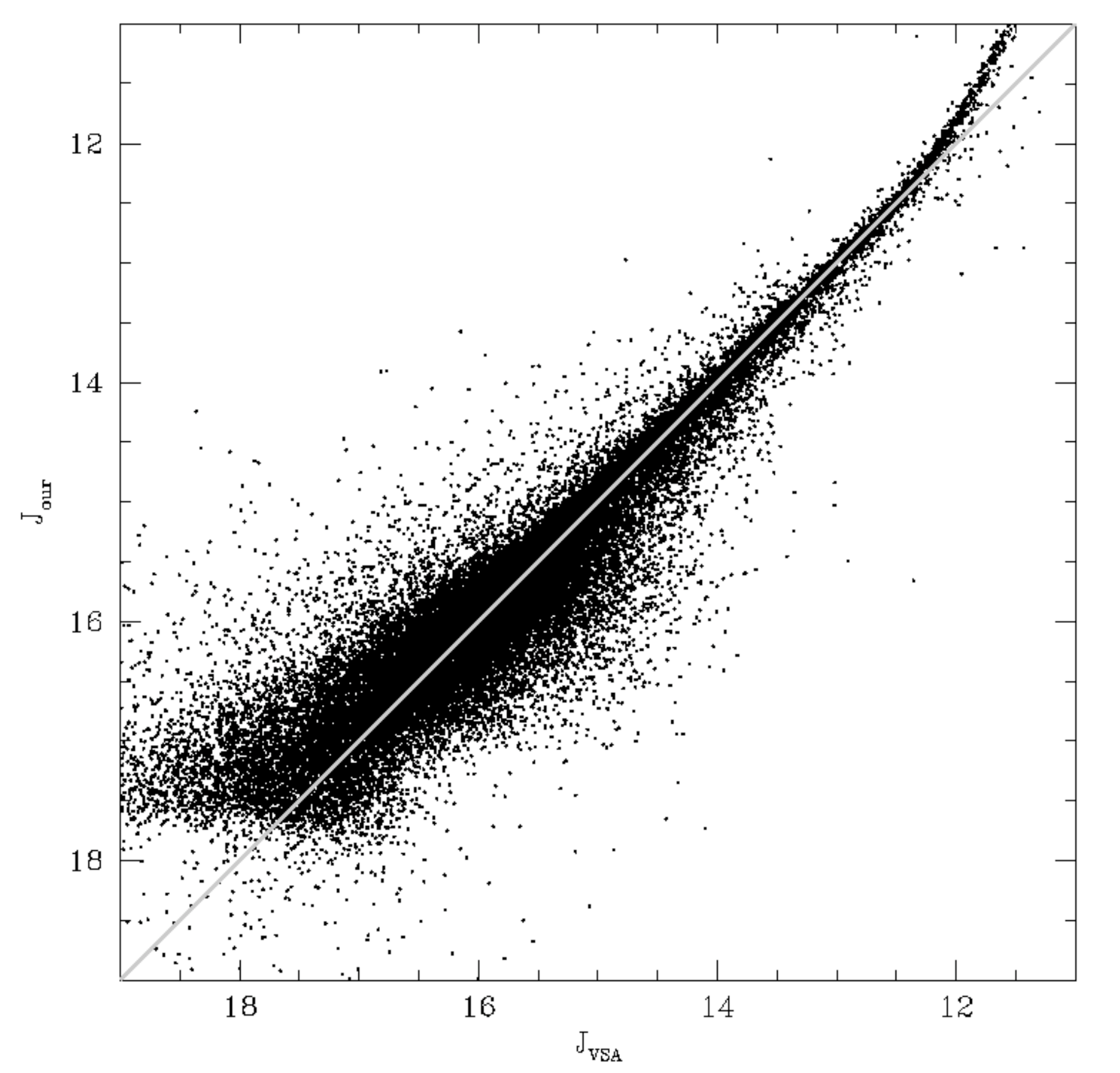}}
\caption{Comparison between our $J$-band magnitude ($J_\mathrm{our}$) and the VSA aperture photometry
catalog ($J_\mathrm{VSA}$), for stars in the stellar field of VVV~CL002. The gray line indicates the
zero-difference between the two photometries ($J_\mathrm{VSA}$=$J_\mathrm{our}$).}
\label{f_casu}
\end{center}
\end{figure}

Stellar photometry was performed with the DAOPHOT\,II and ALLFRAME codes \citep{Stetson94}. After the
construction of the point\--spread function (PSF) for each frame and an initial profile-fitting photometry,
all images of the same object were matched to create a combined frame, where the master list of
star-like sources was created. A PSF fitting of this master list was then performed on each single exposure.
The instrumental magnitudes were transformed into the 2MASS photometric system by means of the 2MASS stars
found in each field. For each frame the 2MASS sources were first identified in our catalog, then the list
was cleaned of objects that in our photometry had a detected source nearer than $2\farcs 2$ and fainter
by less than 3.88 magnitudes. We thus excluded the stars whose 2MASS aperture photometry could have
been contaminated by nearby undetected sources by more than 0.03 magnitudes. We also limited the selection
of calibration stars in the range $J_\mathrm{2MASS}$=11.5-13, $H_\mathrm{2MASS}$=10-11, and
$K_\mathrm{2MASS,s}$=9.5-11, to avoid stars heavily saturated in VVV data and the fainter end of
the 2MASS survey, where systematic errors can be significant (see below). Between 70 and 490 stars met all
criteria in each frame. The calibration equations were in the form
$m_\mathrm{2MASS}-m$=A+B$\times (J-K_\mathrm{s})_\mathrm{2MASS}$, where $m_\mathrm{2MASS}$ and $m$ are the
2MASS and instrumental magnitudes, respectively, and A and B are constants. The color term is in general
negligible, being on the order of few hundredths of magnitude. In Table~\ref{t_calconst} we give the
average coefficients A and B for each pair of offsets collected within a few minutes of each other, and the
rms of the points around the fitted relation. The final calibrated photometry was checked studying the
magnitude difference with respect to the 2MASS catalog. These residuals showed no trend with color, but a
systematic difference was found for the fainter stars in all bands and fields. As an example, the
results of this comparison for VVV~CL002 and its stellar field is shown in Figure~\ref{f_2mas}, but very
similar plots were obtained for the other two cluster candidates as well. The numerous and dispersed
negative values indicate the 2MASS stars contaminated by undetected sources.
The agreement is satisfactory for the brighter stars, but in the faintest $\sim$1.5 magnitudes the bulk of
the distribution curves upward to positive values, indicating that the stars are fainter in the 2MASS
catalog than in our photometry. This feature was found identical even when performing PSF photometry of
VVV data with the DoPhot code \citep{Schechter93}, and when comparing the 2MASS magnitudes with the VVV
aperture photometry downloaded from VSA, whose catalogs agree well with our results. As an example, in
Figure~\ref{f_casu} our $J$ magnitudes in the VVV~CL002 stellar field are compared to the corresponding
quantity in the VSA catalog. The same data that show a systematic difference when compared to 2MASS (upper
panel of Figure~\ref{f_2mas}) do not reveal any trend in this plot, apart from an expected mismatch for
stars in the saturation regime ($J\leq12.5$), where the VSA aperture photometry underestimates the stellar
brightness. We conclude that this systematic offset reflects an intrinsic difference between 2MASS and VVV
data. We believe it is caused by the relative shallowness of the former and its poorer pixel scale, which
leads to an overestimate of the sky background, especially in crowded fields, and consequently to an
overly-faint magnitude estimate for faint stars. A more detailed analysis is beyond the scope of this
paper, and it will be presented in a future study (Mauro et al. 2011, {\it in preparation}).

VVV~CL003 was found in frames acquired with the chips \#12 and \#16 of the VIRCAM camera. The latter chip
is never correctly flat-fielded because of short-timescale variations of its quantum
efficiency\footnote{http://casu.ast.cam.ac.uk/surveys-projects/vista/technical/known-issues}. We found
that the PSF photometry resulting from this chip was unreliable, because of large zero-point variations
on the CCD. We finally decided to exclude these frames, and only the exposures collected with chip \#12,
listed in Table~\ref{t_coord}, were used. Unfortunately, the object fell on the edge of this detector,
where the photometric results are of lower quality, and the cluster center was only $\sim 19\arcsec$ from
the border. Only half of the surrounding field was therefore available for the analysis, and we estimated
by eye that we have studied only about three-fourths of the candidate cluster.

\section{Results}
\label{c_results}

\subsection{Cluster center}
\label{c_cent}

With the final photometry of the three cluster candidates, we revised the location of their centers, first
estimated by eye. This calculation was affected by a smooth background of numerous faint field sources
that shifted the barycenter of the stellar distribution toward the center of the region selected for
the calculation. Cuts in the color-magnitude diagrams did not help, because either the distribution of
cluster and field stars was similar (VVV~CL002 and VVV~CL004, see Figure~\ref{f_cmds}), or the photometric
information was not reliable (VVV~CL003, see \S\ref{c_reduction}). However, if $\vec{\mathrm{r}}$ is the
position of the center of the stellar distribution, and $N$ is the number of stars, we can write
\begin{equation}
\vec{\mathrm{r}}_* N_*=\sum_i \vec{\mathrm{r}}_i=\vec{\mathrm{r}}_\mathrm{c} N_\mathrm{c}+
\vec{\mathrm{r}}_\mathrm{f} N_\mathrm{f},
\label{e_cent1}
\end{equation}
where $\vec{\mathrm{r}}_i$ is the position of the $i$-th star, and the subscripts *, f, and c refer to all,
field, and cluster stars, respectively. After simple algebra, considering that
$N_* =N_\mathrm{c}+N_\mathrm{f}$, and that the center of the distribution of field stars
($\vec{\mathrm{r}}_\mathrm{f}$) coincides with the center of the area included in the calculation
($\vec{\mathrm{r}}_0$), the previous relation yields
\begin{equation}
\vec{\mathrm{r}}_\mathrm{c}=\vec{\mathrm{r}}_0+\frac{\sum_i (\vec{\mathrm{r}}_i-\vec{\mathrm{r}}_0)}{N_*-N_\mathrm{f}}.
\label{e_cent2}
\end{equation}
This formula was used to find the position of the centers of the three objects. The total number of field
stars ($N_\mathrm{f}$) was estimated from the average stellar density in regions farther than 3$\arcmin$
from the initial by-eye center position. To limit the influence of random fluctuations of the stellar
distribution, the calculation was repeated in 360 areas of 3$\arcmin$ radius, whose centers were equally
spaced on a circle of $0\farcm 5$ radius around the initial guess. The results differed by less than
1$\arcsec$ with an rms of only $0\farcs 3$, and they were averaged to obtain the final cluster center,
given in Table~\ref{t_coord}.

To cover a larger field for VVV CL003, the stellar counts were performed including the data collected
with chip \#16. As already noted, these data are not suitable for photometry~-- though they are still good
enough for star count purposes.

\begin{table}[t]
\begin{center}
\caption{Average calibration coefficients for each pair of contiguous offsets.}
\label{t_calconst}
\begin{tabular}{c c c c c c}
\hline
\hline
date & UT & filter & A & B & rms \\
 & & & mag & & mag \\
\hline
\multicolumn{6}{c}{VVV CL002 - chip 12} \\
\hline
2010-04-07 & 7:17:28 & $H$ & 0.17$\pm$0.03 & $-0.013\pm$0.011 & 0.057 \\
2010-04-07 & 7:20:46 & $K_\mathrm{s}$ & $-0.47\pm$0.02 & $-0.060\pm$0.009 & 0.061 \\
2010-04-07 & 7:24:40 & $J$ & 0.64$\pm$0.02 & 0.009$\pm$0.008 & 0.053 \\
2010-04-14 & 9:32:23 & $K_\mathrm{s}$ & $-0.56\pm$0.02 & $-0.028\pm$0.008 & 0.052 \\
\hline
\multicolumn{6}{c}{VVV CL003 - chip 12} \\
\hline
2010-04-07 & 6:52:54 & $H$ & 0.23$\pm$0.02 & $-0.014\pm$0.010 & 0.053 \\
2010-04-07 & 6:56:11 & $K_\mathrm{s}$ & $-0.561\pm$0.016 & $-0.017\pm$0.007 & 0.058 \\
2010-04-07 & 7:01:05 & $J$ & 0.55$\pm$0.02 & 0.039$\pm$0.008 & 0.041 \\
2010-04-14 & 9:06:35 & $K_\mathrm{s}$ & $-0.559\pm$0.014 & $-0.019\pm$0.006 & 0.053 \\
\hline
\multicolumn{6}{c}{VVV CL004 - chip 5} \\
\hline
2010-04-11 & 5:41:06 & $H$ & 0.278$\pm$0.015 & $-0.048\pm$0.010 & 0.033 \\
2010-04-11 & 5:44:13 & $K_\mathrm{s}$ & $-0.554\pm$0.009 & $-0.025\pm$0.006 & 0.036 \\
2010-04-11 & 5:48:40 & $J$ & 0.592$\pm$0.011 & 0.039$\pm$0.008 & 0.032 \\
\hline
\multicolumn{6}{c}{VVV CL004 - chip 9} \\
\hline
2010-04-11 & 5:40:12 & $H$ & 0.230$\pm$0.010 & $-0.014\pm$0.007 & 0.028 \\
2010-04-11 & 5:43:20 & $K_\mathrm{s}$ & $-0.563\pm$0.010 & $-0.014\pm$0.007 & 0.032 \\
2010-04-11 & 5:47:07 & $J$ & 0.622$\pm$0.011 & 0.020$\pm$0.007 & 0.030 \\
\hline
\end{tabular}
\end{center}
\end{table}

\subsection{Significance of the stellar overdensity}
\label{c_signif}

Following \citet{Koposov07}, we can estimate the statistical significance of the stellar overdensities from
the number of stars detected in excess to the local background, whose random fluctuations are assumed to
be Poissonian. We detected 6965 stars in the inner circle of radius 1$\arcmin$ of VVV~CL002, while the field
density level can account only for 5744 objects: an overdensity of stars at the 10.9$\sigma$ level is thus
present in this area. Analogously, 5193 stars are detected for VVV~CL003, against a field expectation of
4313, and the overdensity is significant at the 9$\sigma$ level. The quantity of stars detected in excess of
the field expectation within 1$\arcmin$ from the center of VVV~CL004 is lower ($\sim$400), and the
statistical significance is $\sim$4.5$\sigma$. The 520 square degrees covered by the VVV survey can be
divided into $\sim$600,\,000 circles of radius 1$\arcmin$, therefore the probability of detecting one or
more such overdensity in the whole survey that is caused by random field fluctuations is $\sim$98\%, and
about four similar objects should be expected. Hence, the statistical significance of VVV~CL004 is low,
casting serious doubts on its nature as a true stellar cluster.

\subsection{Radial density profile}
\label{c_radprof}

\begin{figure}
\begin{center}
\resizebox{\hsize}{!}{\includegraphics{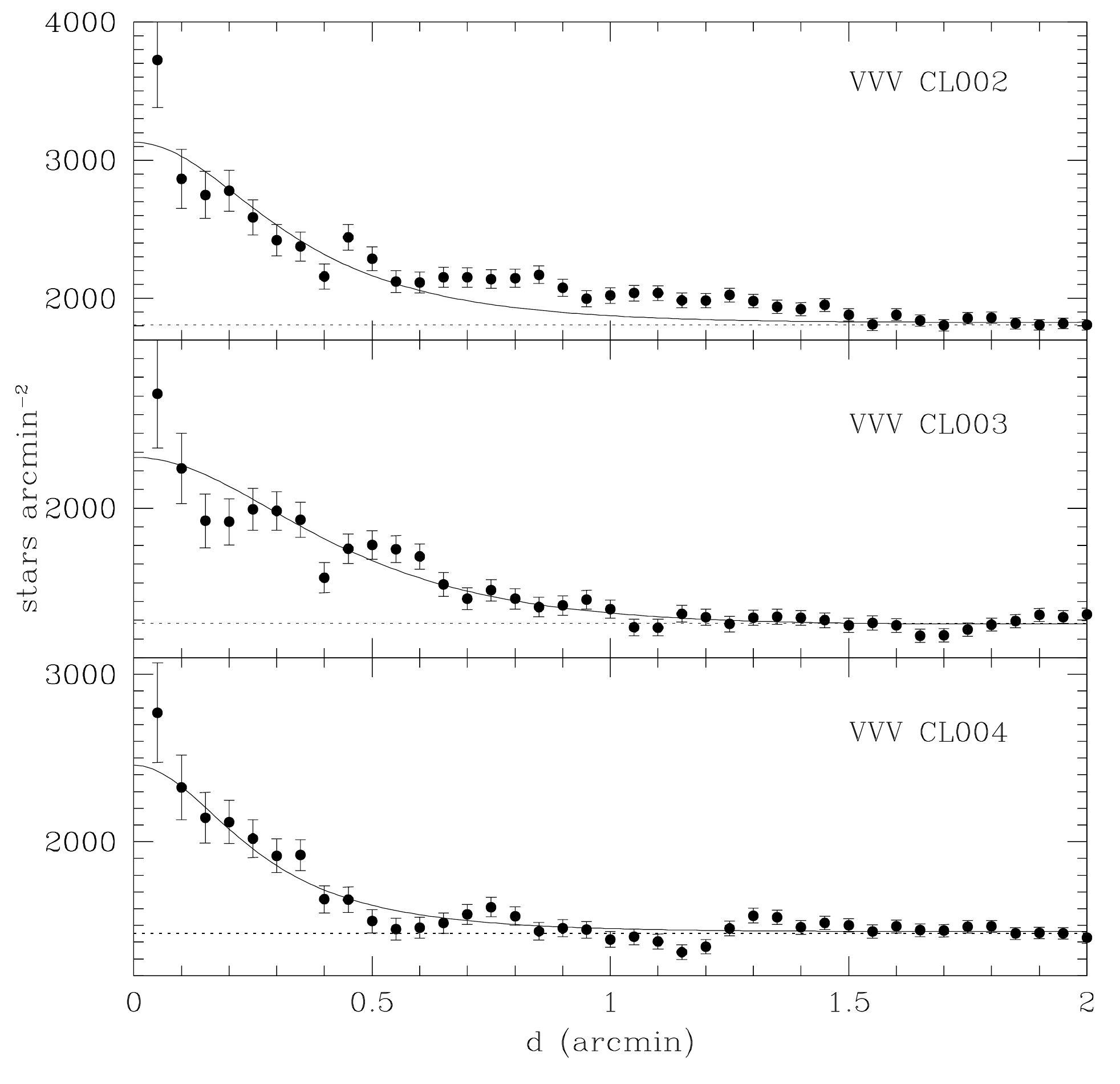}}
\caption{Radial profile of the stellar surface density for the three cluster candidates.
The dotted lines show the estimated field level, while the curve indicates the best-fit King profile.}
\label{f_radprof}
\end{center}
\end{figure}

After the determination of the cluster center, we calculated the stellar density (number of stars per
arcmin$^2$) in concentric rings between the circles of radius $0\farcm 05 \times N$ and
$0\farcm 05 \times N + 0\farcm 1$ (with $N$=0,1,2,3...). The resulting radial profiles of the stellar density
are shown in Figure~\ref{f_radprof}, where the error bars are given by the Poissonian noise. The field density
level, indicated by a dotted line, was assumed as the average value between $1\farcm 8$ and 2$\arcmin$ from the
center, i.e. beyond the clusters' tidal radii determined below. All three objects show a clustering of stars in
the central region. The innermost bin shows a sudden increase of the density in all cases, but this value is
not very reliable because it corresponds to a circle of only $0\farcm 1$ radius. On the other hand, the
possibility of photometric incompleteness in the most crowded region must be taken into account in the
following analysis. If the catalogs were incomplete in the inner part of the cluster candidates, the
statistical significance of the detected stellar overdensity would be underestimated, and the decontaminated
color-magnitude diagrams would be relatively deficient of stars, mainly at fainter magnitudes. However, we
detect between 80 and 100 stars in the inner $0\farcm 1$ of the three clusters, that is 225 times the area
covered by a circle of diameter equal to the FWHM of the PSF ($0\farcs 8$). This indicates that the
three objects are not excessively crowded, and the incompleteness should not be very severe.

\begin{figure}
\begin{center}
\resizebox{\hsize}{!}{\includegraphics{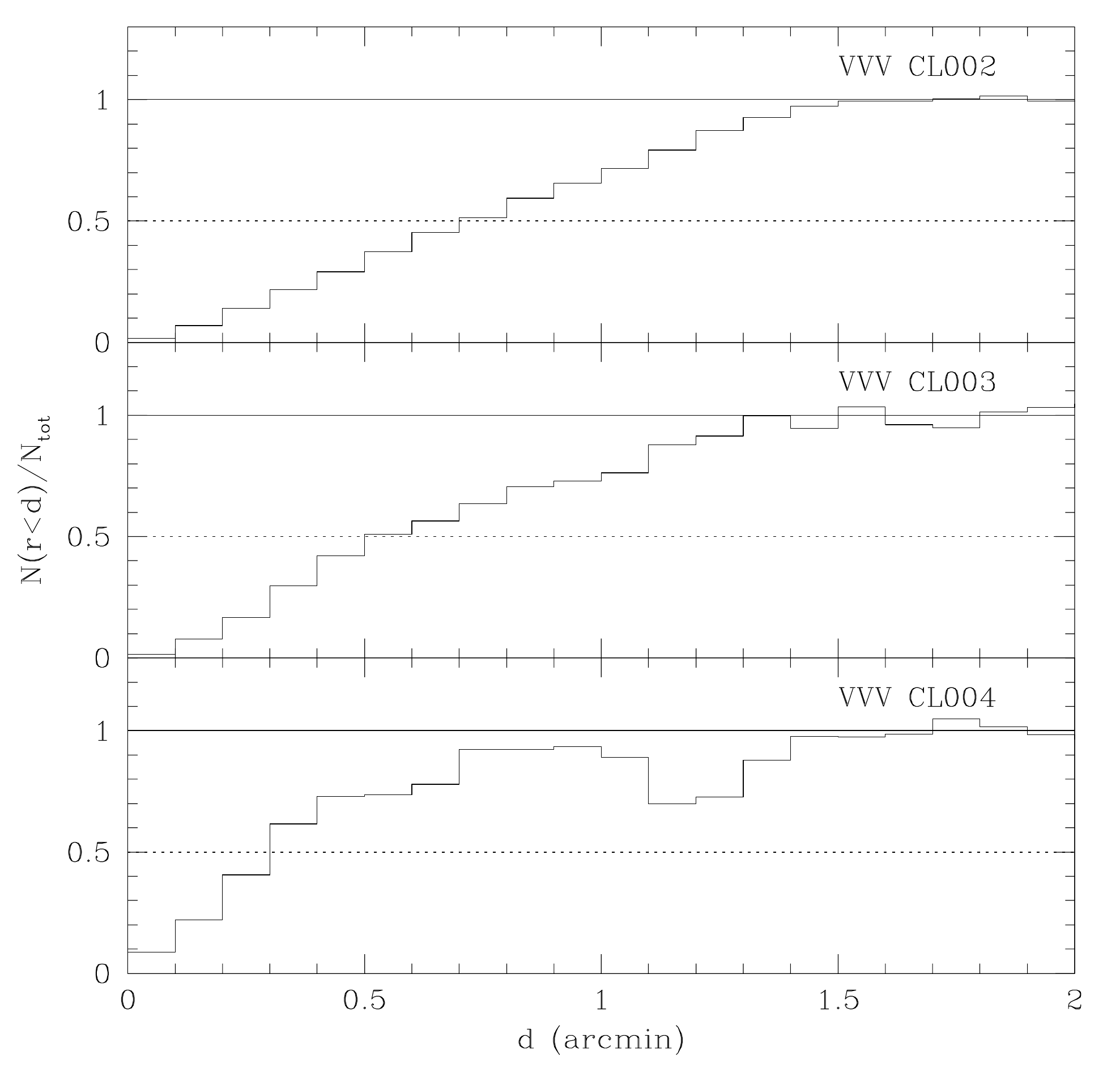}}
\caption{Cumulative fraction of cluster stars as a function of distance from the center. The
field and the half-light levels are indicated with full and dotted lines, respectively.}
\label{f_radfrac}
\end{center}
\end{figure}

The radial profiles of VVV~CL002 and VVV~CL003 are very similar, smoothly falling off at increasing distance
into an extended tail. On the contrary, the density of VVV~CL004 drops more abruptly, but the cluster limit
is not immediately evident because of strong density fluctuations. All these conclusions are confirmed and
better visualized by the radial profiles of the cumulative fraction of stars, presented in
Figure~\ref{f_radfrac}. These were calculated counting the stars detected within a given radius, and
subtracting the number of expected field stars, estimated from the average density between $1\farcm 8$ and
$2\arcmin$ from the center. This range was also used to derive the total number of cluster stars that were
used to normalize the curve. The resulting profiles show that VVV~CL002 extends out to $\sim 1\farcm 6$ from
the center, and its half-light radius, i.e. the radius containing half of the detected stars, is
r$_\mathrm{h}\sim 0 \farcm 75$. The apparent size of VVV~CL003 is probably smaller because it fades into the
field density between $1\farcm 4$ and $1\farcm 6$ from the center, and its half-light radius is
r$_\mathrm{h}\sim 0 \farcm 6$. The angular size of VVV~CL004 is more uncertain, because the field density is
patchy and both the density profile and the cumulative fraction of stars fluctuate strongly. The radius where
the density merges into the field continuum is unclear, probably comprised between 1$\arcmin$ and $1\farcm 7$,
while r$_\mathrm{h}\sim 0 \farcm 35$. The radial density profile of the three objects was fitted with a
\citet{King66} profile, and the results are overplotted on the observational data in Figure~\ref{f_radprof}.
For VVV~CL003 this procedure returned a very good fit of the data, with a tidal radius
$r_\mathrm{t}=1\farcm 8$ and a central concentration $c=\log{(r_\mathrm{t}/r_\mathrm{c})}=0.56$, where
$r_\mathrm{c}$ is the core radius. The density profile of VVV~CL004 is also satisfactorily reproduced by the
fitted function, and the cluster parameters derived in this way are $r_\mathrm{t}=1\farcm 7$ and $c=0.75$. On
the other hand, the routine failed to return physically reasonable parameters for VVV~CL002, because the
density between $0\farcm 5$ and $1\farcm 5$ is too high, causing $r_\mathrm{t}$ to diverge. A shallower slope
of the density profile indicates the presence of tidal tails \citep{Kuepper10a}, and the cluster may be
losing stars along its orbit. We therefore estimated $r_\mathrm{t}=1\farcm 8$ from the point where
N(r$<$d)/N$_\mathrm{tot}$=1 in Figure~\ref{f_radfrac}, then a King profile of fixed tidal radius was fitted
to find $c=0.65$. Similar problems were found with a \citet{Wilson75} profile, calculated from the
energy distribution function through the code of \citet{Sollima09}: fitting both the central decay
($r\leq 0\farcm5$) and the enhanced tail was impossible. In this case the fitting routine converged, but to a
very high tidal radius ($c$=0.68, $r_\mathrm{t}=4\farcm 8$), and the resulting fit was unsatisfactory. We
found that a EFF power-law \citep{Elson87} fits the observed profile well, with central density $\mu_0$=5630
stars, normalization radius $a=0\farcs 16$, and power $-$0.086 ($\gamma$=0.17). The fitting routine likely
underestimates the uncertainties associated with its output, because it does not take into account any source
of systematic error, mainly the definition of the background level. We consider that the width of the bins of
the density profile ($0\farcm 1$) is a better estimate of the errors on the core, half-light, and tidal radii.
Indeed, we found that these quantities can vary by up to $0\farcm 1$ from the given values when different but
still reasonable assumptions on the background level were adopted.

\subsection{Color-magnitude diagrams}
\label{c_cmds}

\begin{figure}
\begin{center}
\includegraphics[width=7.7cm]{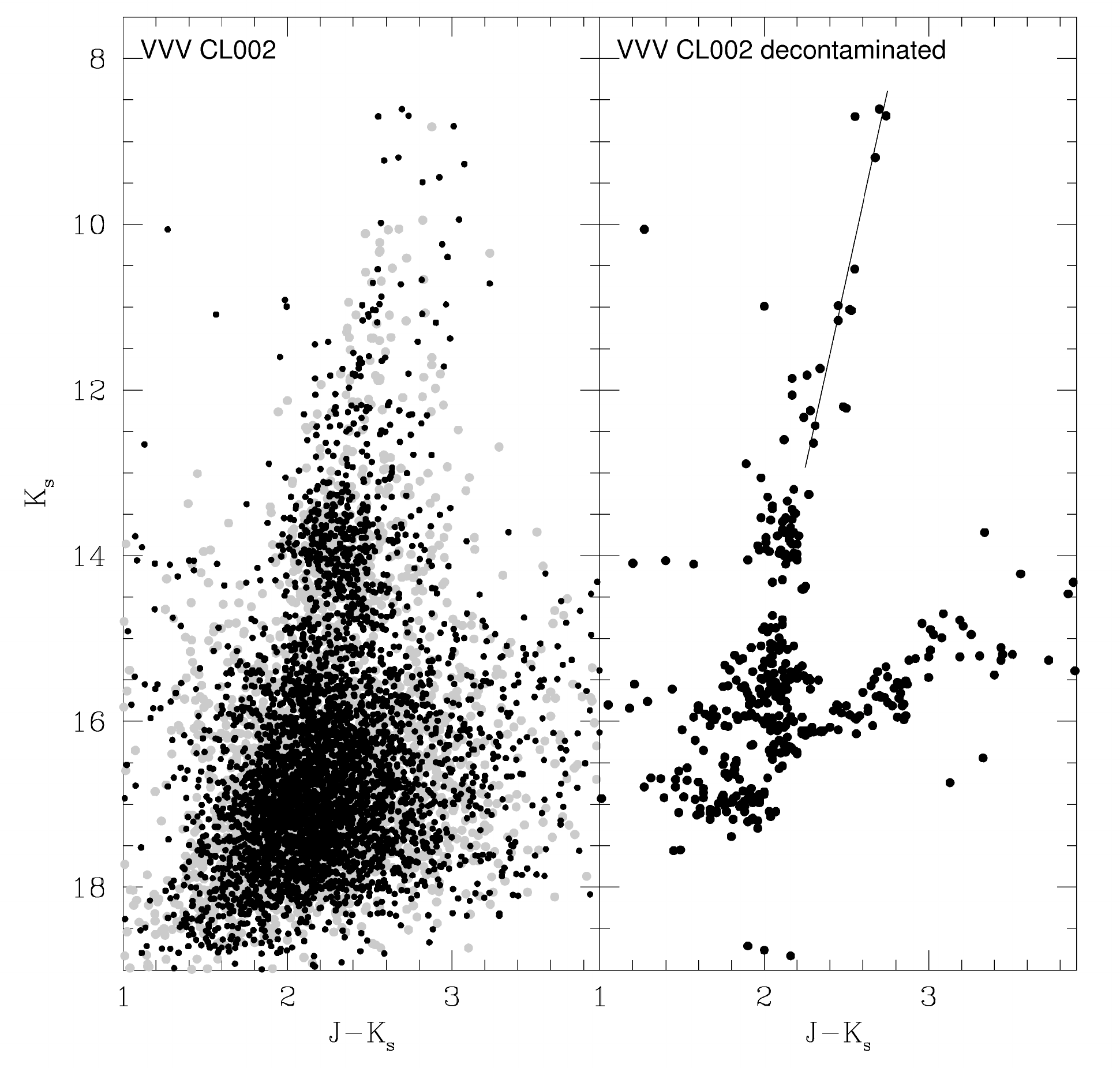}
\includegraphics[width=7.7cm]{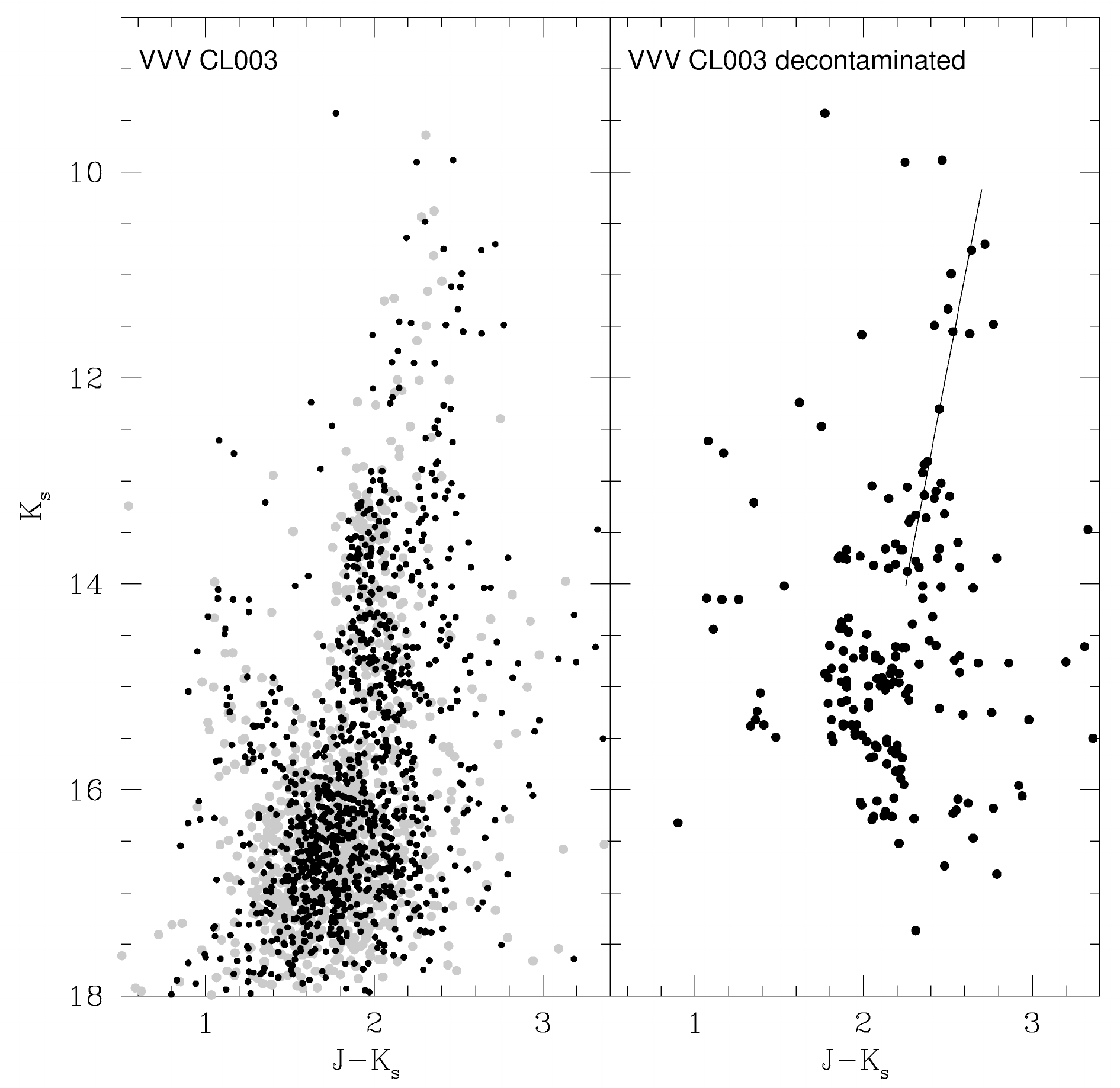}
\includegraphics[width=7.7cm]{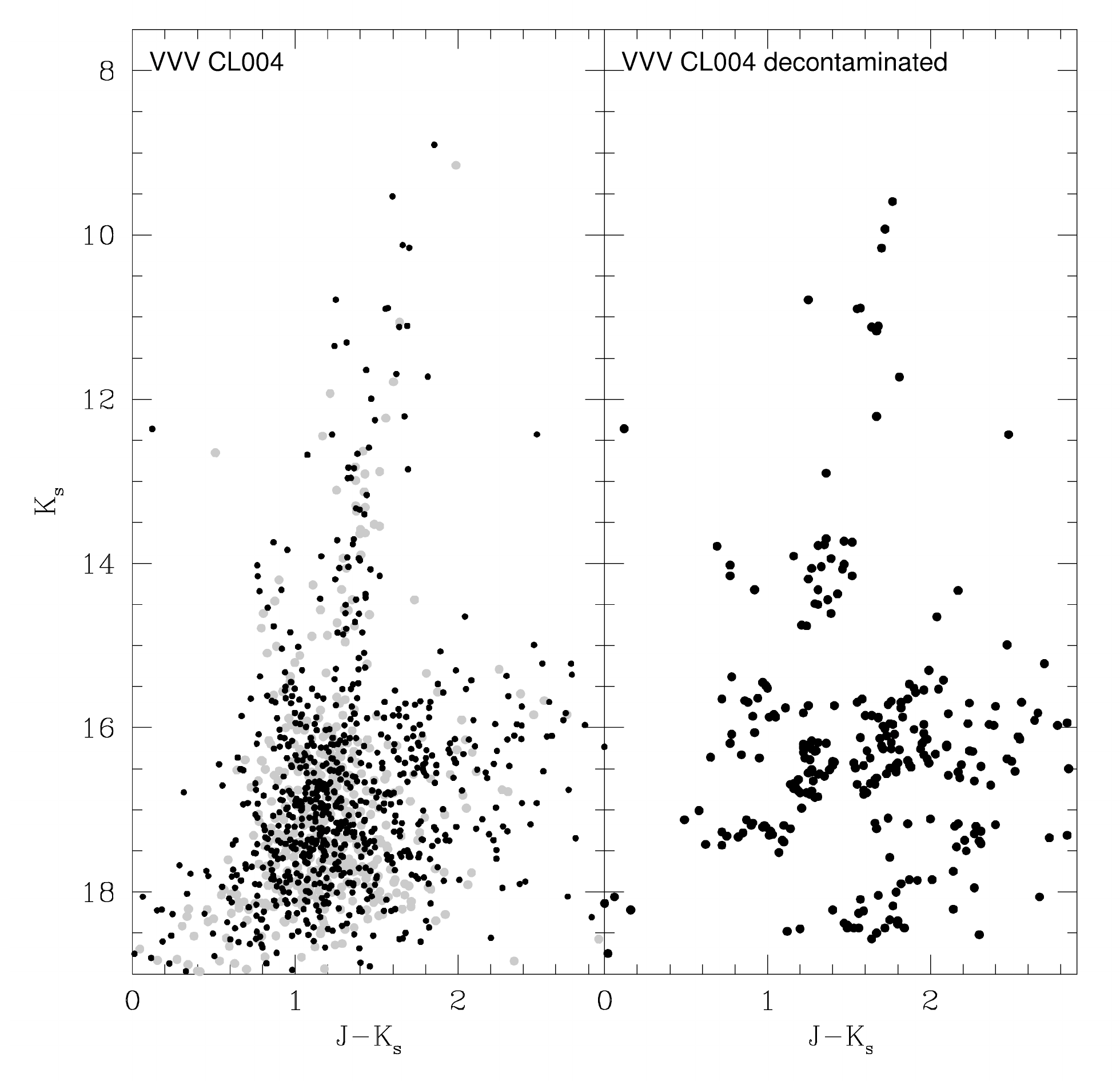}
\caption{{\it Left panels}: CMDs of the three cluster candidates (black dots) and
their comparison fields (gray dots). {\it Right panels}: CMDs after statistical
subtraction of the field contaminants.}
\label{f_cmds}
\end{center}
\end{figure}

\begin{figure}
\begin{center}
\resizebox{\hsize}{!}{\includegraphics{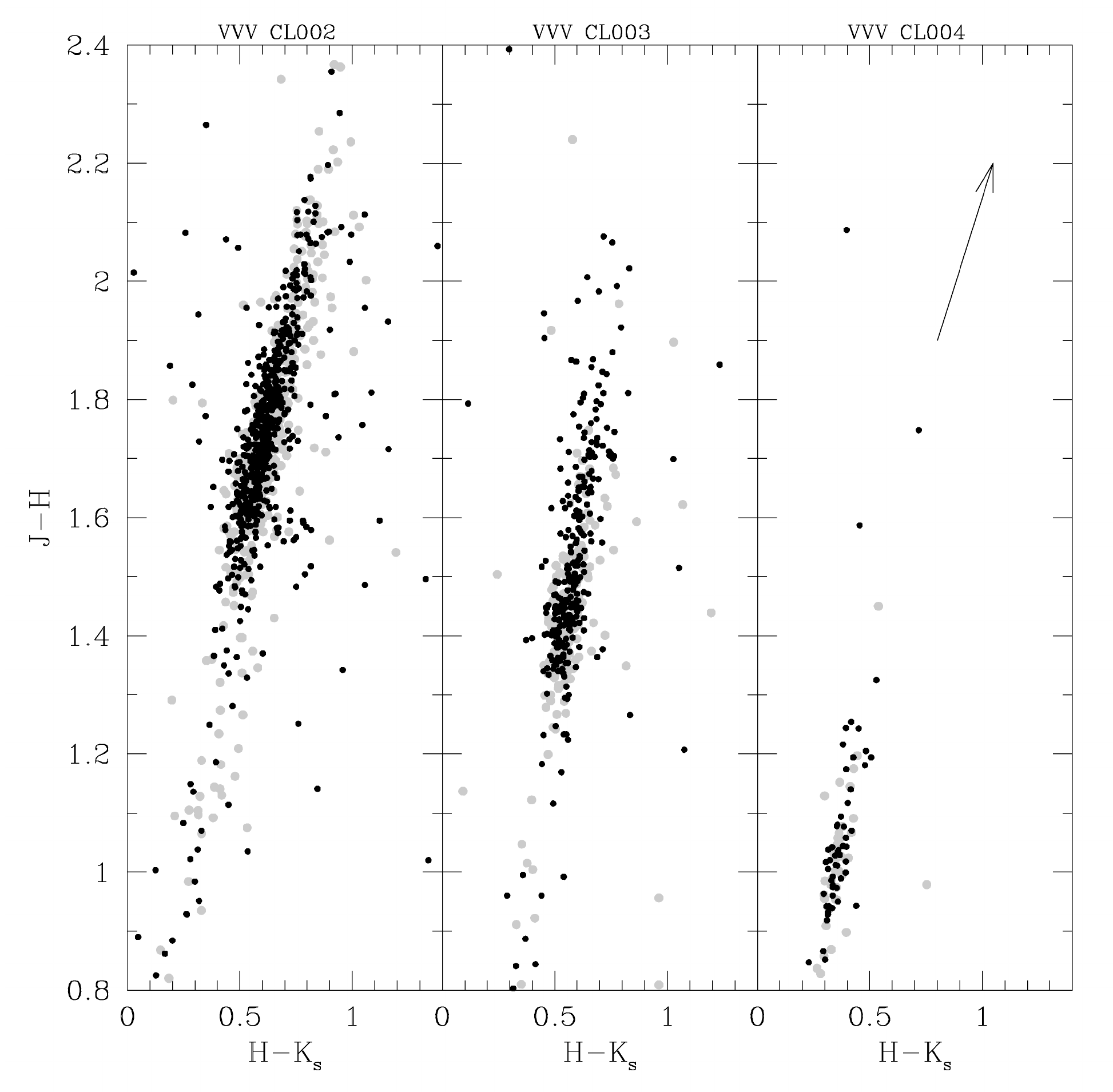}}
\caption{Color-color diagrams of the three cluster candidates (black dots) and their comparison
fields (gray dots). The plot was restricted to stars brighter than K$_\mathrm{s}$=15. The arrow in
the right panel shows the direction of the reddening vector.}
\label{f_2col}
\end{center}
\end{figure}

The color-magnitude diagrams (CMDs) of the three objects are shown in the left panels of Figure~\ref{f_cmds},
where the cluster stars are overplotted on the photometry of a nearby comparison stellar field. The cluster
area of VVV~CL002 and VVV~CL004 was limited to a circle of radius equal to the half-light radius to
minimize the background contamination. The comparison field was an annulus of the same center and area as the
cluster, and inner radius of 1$\farcm$8, i.e. just outside the cluster tidal radius. A similar definition of
cluster and field areas for VVV~CL003 was complicated because its photometric catalog lacked the
required symmetry about the center. We adopted as cluster region a circle of radius $0\farcm 5$, but part of
this area fell outside the CCD. The definition of the comparison field was also complicated because VVV~CL003
is found in a region affected by a very strong reddening gradient. Indeed, the interstellar extinction varies
by up to $\Delta$E($B-V)\approx$2 over the 10$\arcmin$ of the chip, as confirmed by inspection of the
\citet{Schlegel98} maps. We noticed, however, that the CMD of the cluster area presents two well-defined red
giant branches (RGBs): a narrow and well-populated one at ($J-K_\mathrm{s})\approx$2, and a broader RGB about
0.5~magnitudes redder. The bluer sequence was found in all studied areas, always very similar in terms of
quantity of stars, width, and magnitude of the red clump, although shifted to different colors. The other RGB
component, on the other hand, was not observed outside $\sim 1\arcmin$ from the cluster center. We therefore
concluded that the bluer RGB represents the field population, and we searched for a region in the vicinity of
the cluster where the color of this feature coincided with that observed in the cluster area. The adopted
region had the same shape and dimension as the defined cluster area, but offset $1\arcmin$ toward the
northwest, i.e. perpendicularly to the Galactic plane toward increasing Galactic latitude. Part of the object
could still fall in this area, but its density should be very low at this distance. In Figure~\ref{f_2col} we
show the color-color diagrams of the three objects and their comparison fields. This plot was restricted to
stars brighter than K$_\mathrm{s}$=15 to highlight the color differences at the RGB level, because the
increased photometric errors blur them for fainter stars.

As already noted, the CMD of VVV~CL003 shows two distinct RGB sequences, and the redder one is not
observed in the surrounding stellar field. In the comparison field, some residual cluster stars are
also visible, not observed at longer distances from its center. On the other hand, the CMDs of
VVV~CL002 and VVV~CL004 show no immediately clear difference with their comparison field, except
for the enhanced number of stars. Further inspection reveals that the cluster RGB of VVV~CL002 is
on average bluer than the field population. This is visible even in the color-color diagrams, where
cluster stars of VVV~CL002 are offset to slightly bluer colors with respect to the field population.
We found no color gradient that can explain this behavior in the whole
$\sim 10\arcmin \times 10\arcmin$ area covered by our photometry, although the reddening pattern is
probably variable on a small scale, as indicated by some highly reddened stars found in both cluster
and comparison fields. VVV~CL004 is less clear than the other two objects. Its color-color diagram
contains more red stars than the surrounding field, but this is a simple consequence of the higher
number of stars in the upper (redder) RGB of the cluster area. Indeed, although some bright
(K$_\mathrm{s}$=11-13) red stars without field counterparts are observed, the CMD of VVV~CL004 shows
no clear peculiarity with respect to the field, except for being much richer in brighter stars,
as well as in faint red stars.

\subsection{Statistical field decontamination}
\label{c_decont}

The CMDs of Figure~\ref{f_cmds} are strongly contaminated by field stars, and so we applied a
statistical procedure to isolate the cluster CMDs. The code written for this purpose was based on
the method of \citet{Gallart03}. In brief, a metric was defined in the
$K_\mathrm{s}$-$(J-K_\mathrm{s})$ plane as
\begin{equation}
d=\sqrt{(K_\mathrm{s,a}-K_\mathrm{s,b})^2+(k\times ((J-K_\mathrm{s})_a-(J-K_\mathrm{s})_b))^2},
\label{e_metric}
\end{equation}
where $k$ is an arbitrary coefficient weighting a difference in color with respect to a difference
in $K_\mathrm{s}$. The code scanned the list of stars in the comparison field and found the nearest
star (smaller $d$) in the CMD of the cluster area for each object. This source was rejected as a
contaminating field star if its distance $d$ was smaller than an arbitrary threshold
$d_\mathrm{max}$, otherwise the star in the comparison area was flagged as an ``object without a
counterpart". In the first case, the list of stars in the cluster area was updated after the
rejection before proceeding to the next entry in the list.

The choice of the parameters $d_\mathrm{max}$ and $k$ is not completely free of arbitrariness.
\citet{Gallart03} used $d_\mathrm{max}$=0.3 magnitudes, and we found this is a good choice in our case
also, because it prevented the association between two stars with a very different position in the CMD.
The same authors adopted $k$=7 in their optical photometry, which gives a dominant role to an offset
in color. Their choice was dictated by the need to remove foreground contaminants distributed in a wide
range of distances, with only limited variations of reddening. In our case, though, the patchy and
locally variable structure of the reddening makes a color offset between cluster area and comparison
field more likely than a difference in magnitude. Moreover, ($J-K_\mathrm{s}$) is more sensitive than
the $K_\mathrm{s}$ band to interstellar absorption, and even a different distance can affect the
stellar color in our case, because we are observing through the dense clouds of the inner Galaxy. In
conclusion, a much lower value of $k$ should be appropriate in our study. Indeed, we explored the
effects of this choice through repeated trials, searching for the value that returned a clearer CMD
and a smaller fraction of comparison field stars that were without counterpart. We found that both the
residual contamination in the product CMDs and the quantity of unmatched field stars decreased
with $k$, and $k\leq$2 was clearly preferred. Only negligible changes were observed when the scaling
factor was varied in the range $k$=1--2, and we finally adopted $k$=1.6. The unmatched field stars
thus decreased to $\sim$2\% for VVV~CL004, which is below the Poissonian noise, while it was
$\sim$10\% for VVV~CL002 and VVV~CL003, i.e. about 3--4 times the Poissonian noise. However, the
decontaminated CMDs of VVV~CL002 and VVV~CL003 contain 453 and 211 stars, respectively, which matches
the expectation of 454 and 250~objects well, that was a result of the stellar counts of
Section~\ref{c_radprof}. Hence, the decontamination procedure worked satisfactorily.

In the decontamination procedure, different choices of the cluster and field areas necessarily led
to different sets of removed stars. However, all features observed in the resulting CMDs, such
as sequences and clumps, were observed in all cases when the cluster center was unchanged. The
innermost regions of the three objects were excluded from the calculation, because our photometric
catalogs could be incomplete in the most crowded regions. In general, the cleaning procedure left
some residual field contamination, that decreased when choosing $i$) smaller areas, which consequently
reduced the number of contaminating stars to be removed, and $ii$) a comparison field closer
to the cluster center, which minimized the effects of a different local reddening. In both cases,
the lower level of residual contamination came at the cost of a poorer derived CMD, because cluster
members were also excluded. Nevertheless, these choices are justified here, because our aim is to
unveil the behavior of the cluster CMD, while the true number of stars in the different CMD branches
(i.e. the detailed luminosity function) is of lesser importance. Finally, the cluster area
adopted in the decontamination procedure of VVV~CL002 and VVV~CL004 was a circle of radius
$0\farcm 5$ and $0\farcm 3$, respectively, slightly smaller than their half-light radii. The
inner $0\farcm 1$ from the center was also excluded. The comparison fields were annuli of equal area
as the cluster region, with inner radius of $0\farcm 9$ in the case of VVV~CL002 and $0\farcm 7$
for VVV~CL004. The cluster density is very low at these distances, and the loss of cluster
stars in the resulting CMDs should be limited. Yet the areas of VVV~CL003 and its
comparison field were defined as in \S\ref{c_cmds} (two circles of $0\farcm 5$ radius, offset by
$1\arcmin$) because the strong underlying reddening gradient and the position of the cluster at the
edge of the CCD gave no freedom for exploring different choices. Even for VVV~CL003 the inner region
with $r\leq0\farcm 1$ from the center was excluded.

The decontaminated CMDs of the three cluster candidates are shown in the right panels of
Figure~\ref{f_cmds}. In all cases a clear evolutionary sequence of stars is visible, with a clumpy
behavior caused by some underpopulated regions, and many highly scattered stars mainly at the fainter
end of the diagrams. These features are probably an artifact of the decontamination procedure: when
the cluster and field CMDs are very similar and the sample is statistically small, stars in crowded
regions of the CMD are more easily matched with a field counterpart than the objects deviating
because of high random error or higher differential reddening. As a consequence, the algorithm could
over-subtract stars along the stellar sequence and under-subtract them in the wings of the color
distribution, while still subtracting the expected amount of field contaminants, as indicated by the
quantity of cluster and unmatched field stars given before.

\subsection{Reddening-free indices}
\label{c_marcindex}

\begin{figure}
\begin{center}
\resizebox{\hsize}{!}{\includegraphics{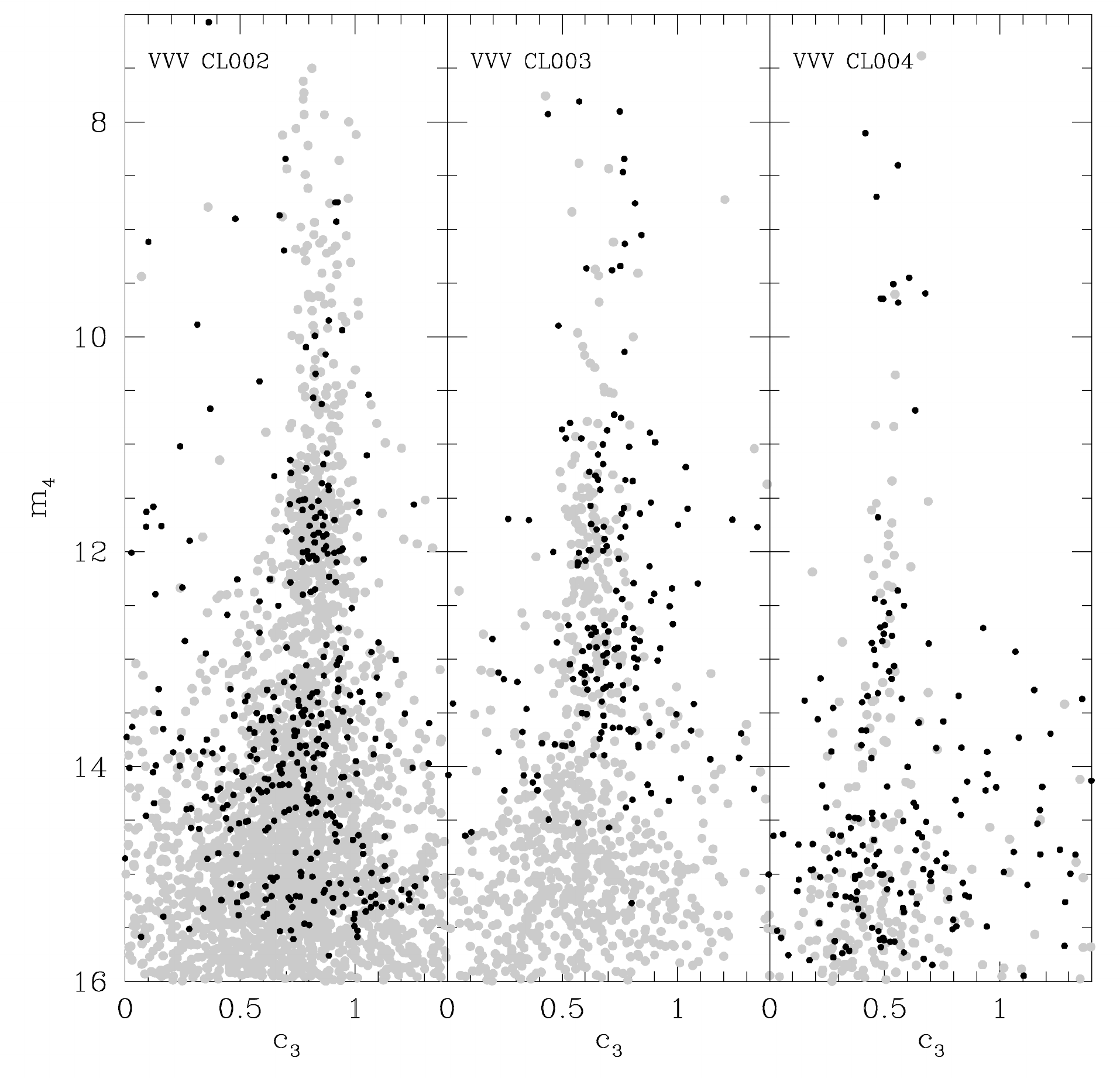}}
\caption{Reddening-free CMDs of the three field decontaminated cluster candidates (black
dots), overplotted on their comparison fields (gray dots).}
\label{f_marcindex}
\end{center}
\end{figure}

In the presence of non-uniform, spatially variable reddening, it is instructive to investigate
the CMDs where the color and magnitude are substituted by reddening-free indices. In
Figure~\ref{f_marcindex} we show the decontaminated CMDs of the three objects overplotted on
their comparison field, with the indices $c_3=(J-H)-1.47\times(H-K_\mathrm{s})$ and
$m_4=K_\mathrm{s}-1.22\times(J-H)$ in place of the color and magnitude. These quantities were
defined by \citet{Catelan11} assuming the \citet{Cardelli89} relation between extinction in
different bands, and R=A$_\mathrm{V}$/E($B-V$)=3.09 \citep{Rieke85}. We also experimented
with analogous indices derived assuming the near-IR extintion law for the central Galactic
regions \citep{Nishiyama09}, but they led to the same general conclusions. Unfortunately, the
error on $c_3$ and $m_4$ blurs the resulting diagram, because it is about three times larger than
that in ($J-K_\mathrm{s}$) and $K_\mathrm{s}$ respectively, owing to propagation of errors. The
average uncertainty on these indices is about 0.1 mag down to $m_4$=10, 0.15 mag at $m_4$=12, and
it increases exponentially at fainter magnitudes. Thus most of the information is erased.
Moreover, while zeroing the effects of reddening, the indices also strongly limit the effect of
other parameters, such as age or metallicity. For example, we verified through synthetic isochrone
experiments that the RGBs of two stellar populations of the same age but different metallicity,
with $[\mathrm{Fe/H}]$=0.0 and $-$1.5, would form two nearly parallel sequences separated by less
than $\Delta{c_3}\approx$0.2. This, combined with the large errors, makes two different
populations difficult to distinguish. Nevertheless, Figure~\ref{f_marcindex} shows that the
cluster sequence is redder than the field in VVV~CL003. For VVV~CL002 the difference is less
evident, but both a linear and a quadratic fit of the upper RGB ($m_4\leq$12.5) are between 0.01
and 0.04~magnitudes bluer in the cluster than in the field. Then, the color differences between
these objects and their surrounding fields are not caused by differential reddening alone.

Figure~\ref{f_marcindex} does not reveal much about VVV~CL004, because field and cluster
sequences already overlapped in the ($J-K_\mathrm{s}$)-$K_\mathrm{s}$ plane. It can be noted,
however, that some red stars in the upper RGB ($K_\mathrm{s}=11-13$, see Figure~\ref{f_cmds})
agree with the field locus in the $c_3$-$m_4$ diagram, indicating that some differential
reddening is present even in the direction of this object.

\subsection{Proper motions}
\label{c_pm}

\begin{figure}
\begin{center}
\includegraphics[width=7.7cm]{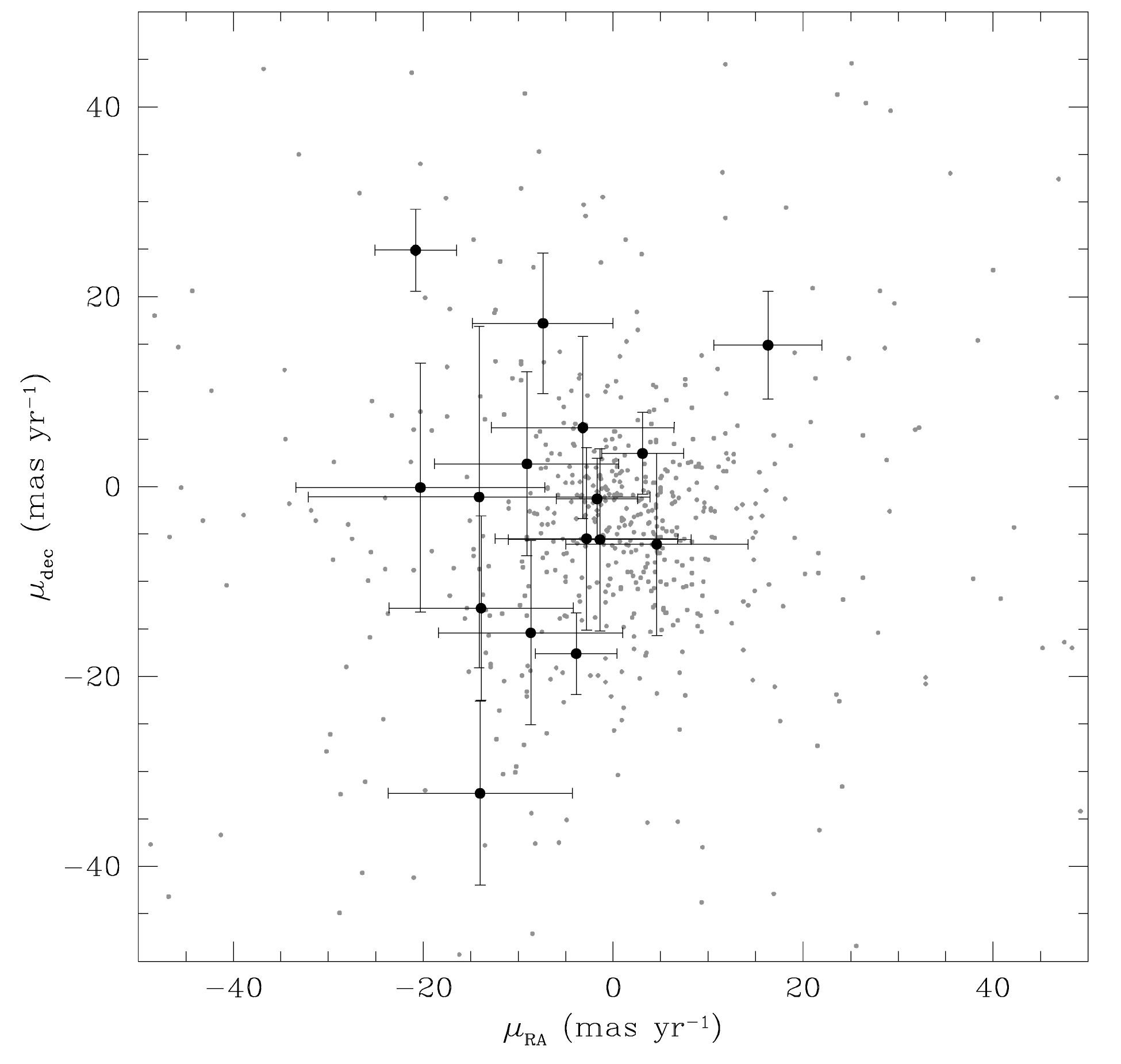}
\includegraphics[width=7.7cm]{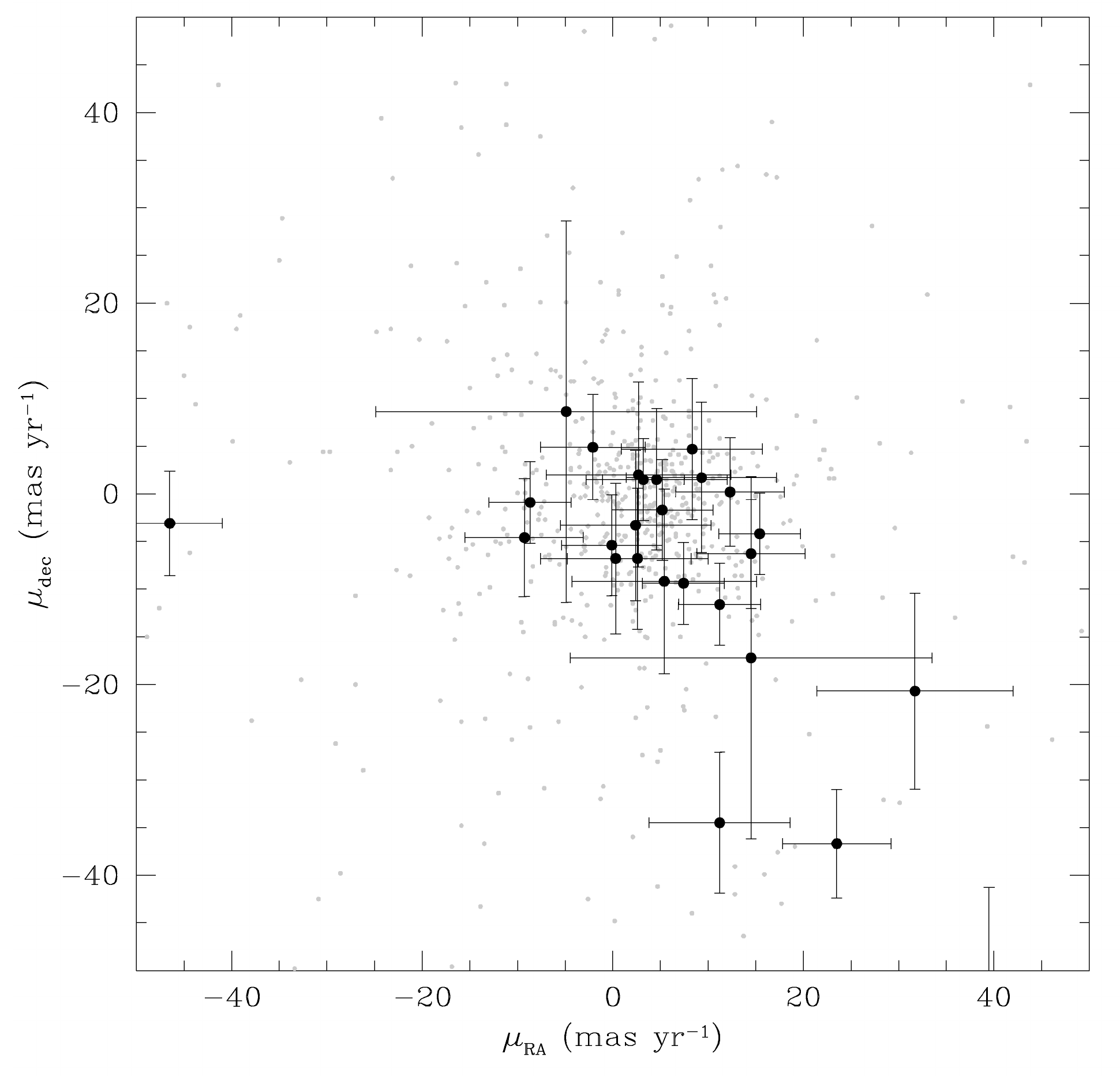}
\includegraphics[width=7.7cm]{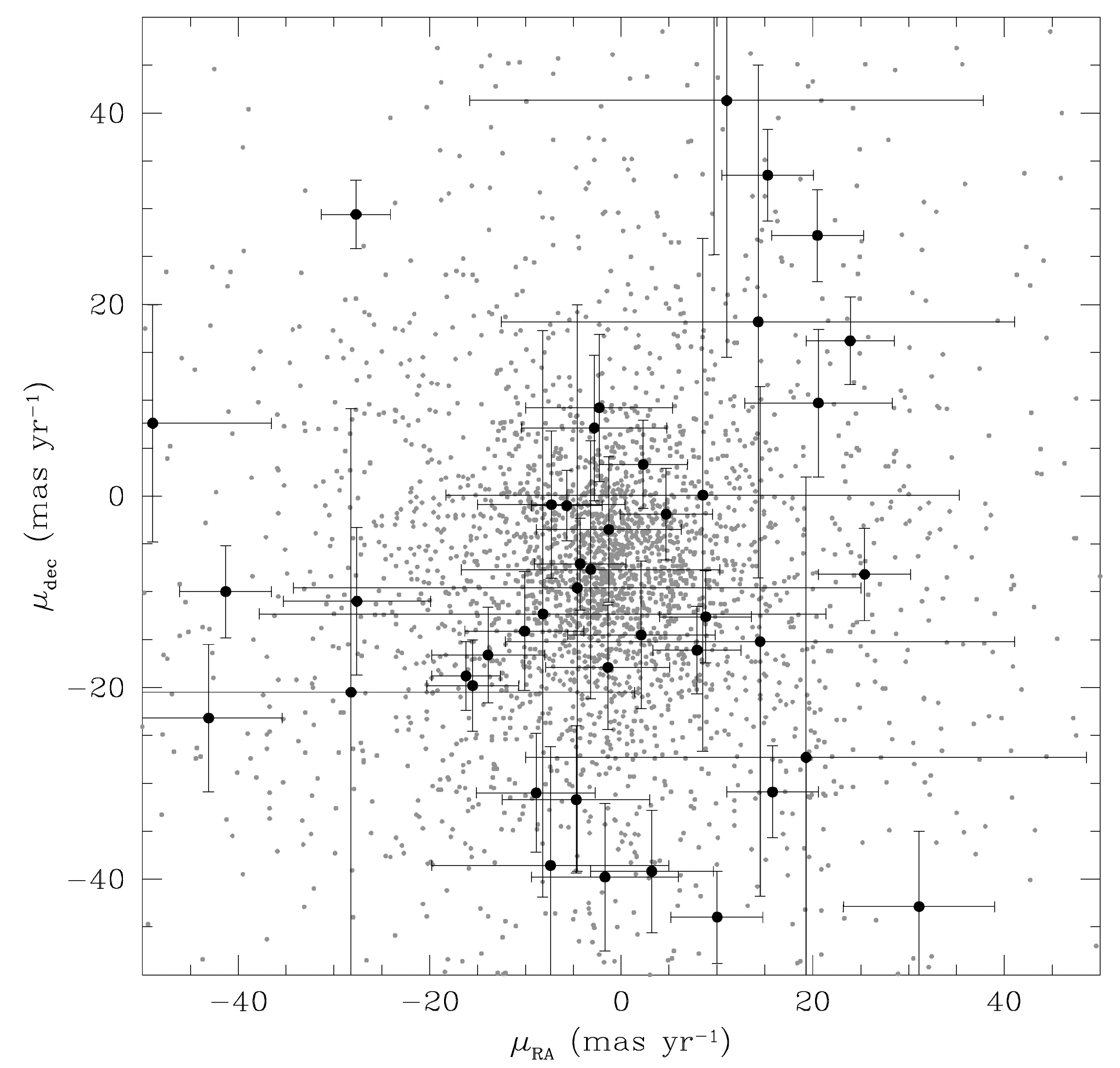}
\caption{Distribution of proper motions in the cluster area (full dots with error bars) and
in the comparison field (small gray dots). {\it Upper panel}: VVV~CL002;
{\it Upper panel}: VVV~CL003; {\it Lower panel}: VVV~CL004.}
\label{f_pm2}
\end{center}
\end{figure}

The information about proper motion of stars in the field of an observed stellar overdensity
is a very powerful tool to clarify its nature, especially if combined with the photometric data
\citep[e.g.,][]{Carraro05,Pavani07}. Indeed, a study of new cluster candidates
without an analysis of the proper motions would seem incomplete. Unfortunately, it
provides no real information for our three objects, because available data are scarce, and at
the estimated distances ($\sim$10~kpc, see \S\ref{c_paramD2}-\S\ref{c_paramD3}) a too high
precision is required to detected an appreciable difference with the field population. For
example, the UCAC3 \citep{Zacharias10} and NOMAD \citep{Zacharias04} catalogs contain only 2
and 11~stars, respectively, in the inner 1$\arcmin$ of VVV~CL002.

The proper motions of stars in the direction of VVV~CL003 and VVV~CL004 and in their surrounding
fields drawn from the PPMXL catalog \citep{Roeser10} are compared in the middle and lower panel
of Figure~\ref{f_pm2}. The cluster stars were taken within $0\farcm 8$ (VVV~CL003) and
$0\farcm 5$ (VVV~CL004) from the center, and the comparison field was defined as an annulus
between 2$\arcmin$ and 5$\arcmin$. Clearly, this comparison is inconclusive, and the average value
in the cluster and field area do not differ by more than 1~mas~yr$^{-1}$ in both components.
According to our stellar counts, only 11\% and 13\% of the stars in the selected region should be
members of VVV~CL003 and VVV~CL004, respectively. Attempting to isolate cluster members of
VVV~CL003 through the 2MASS photometry given in the PPMXL catalog, only three stars compatible
with the cluster sequence can be identified.

The proper motions of stars within $0\farcm 8$ from the center of VVV~CL002 are offset with
respect to the comparison field, as shown in the upper panel of Figure~\ref{f_pm2}. Only 4.4
cluster members ($\sim$20\%) are expected among the 22 entries of the PPMXL catalog, and its very
hard to assess the significance of the observed behavior. As an exercise, we will check if this
offset is compatible with the detection of the peculiar movement of VVV~CL002. Assuming the
escape velocity from the Galactic bulge \citep[$\sim 600~\mathrm{km~s}^{-1}$,][]{Lamb96} as an
upper limit for the tangential velocity, the proper motion of stars at 7.5 kpc from the Sun should
not exceed 17 mas~yr$^{-1}$. We will therefore restrict the calculation to the stars with
$\vert \mu\vert\leq$20~mas yr$^{-1}$ because, considering the average error ($\sim$8~mas yr$^{-1}$),
the cluster stars should not be found outside this range. The average values are
($\bar{\mu_{\alpha}},\bar{\mu_{\delta}}$)=($-$2.8,$-$0.5) and (0.8,$-$3.0) for the cluster and
field stars, respectively. In the cluster area we can write
$N_\mathrm{tot}\bar{\mu}=N_\mathrm{f}\bar{\mu_\mathrm{f}}+N_\mathrm{c}\bar{\mu_\mathrm{c}}$ for
both components of proper motion, where the subscripts $f$ and $c$ denote the field and cluster
members, respectively, and $N$ is the number of stars. We have $N_\mathrm{tot}$=12, and assuming
$N_\mathrm{c}$=5, $N_\mathrm{f}$=7, we obtain ($\bar{\mu_{\alpha}},\bar{\mu_{\delta}}$)=($-$7.7,3.0)
for cluster stars, which translates into a heliocentric tangential velocity of $\sim$300 km s$^{-1}$
at 7.5~kpc. This is somewhat high, but an acceptable value. The observed behavior of proper motions
in the field of VVV~CL002 can therefore reflect the presence of a comoving group of stars.

\subsection{The nature of the cluster candidates}
\label{c_nature}

Although the observed radial density profiles are well\--represented by the King law, which is typical of
stellar clusters, and the stellar overdensity associated with VVV~CL002 and VVV~CL003 is statistically
highly significant, we will not rely on stellar counts alone to conclude about the nature of the three
cluster candidates, because they can sometimes be misleading
\citep[see, for example,][]{Villanova04,Moni09}. Unfortunately, the proper motions do not provide useful
information about this.

The case of VVV~CL003 is straightforward, because the cluster sequence in the CMD is clearly distinct
from the field one, hence this object cannot be only a fluctuation of the field stellar density. The
redder putative cluster sequence cannot be ascribed to a local fluctuation of reddening, because in this
case the extinction would be higher: a lack of stars would thus be expected instead of the observed
significantly higher stellar counts. The reddening-free CMD also supports this conclusion.

\begin{figure}
\begin{center}
\includegraphics[width=9cm]{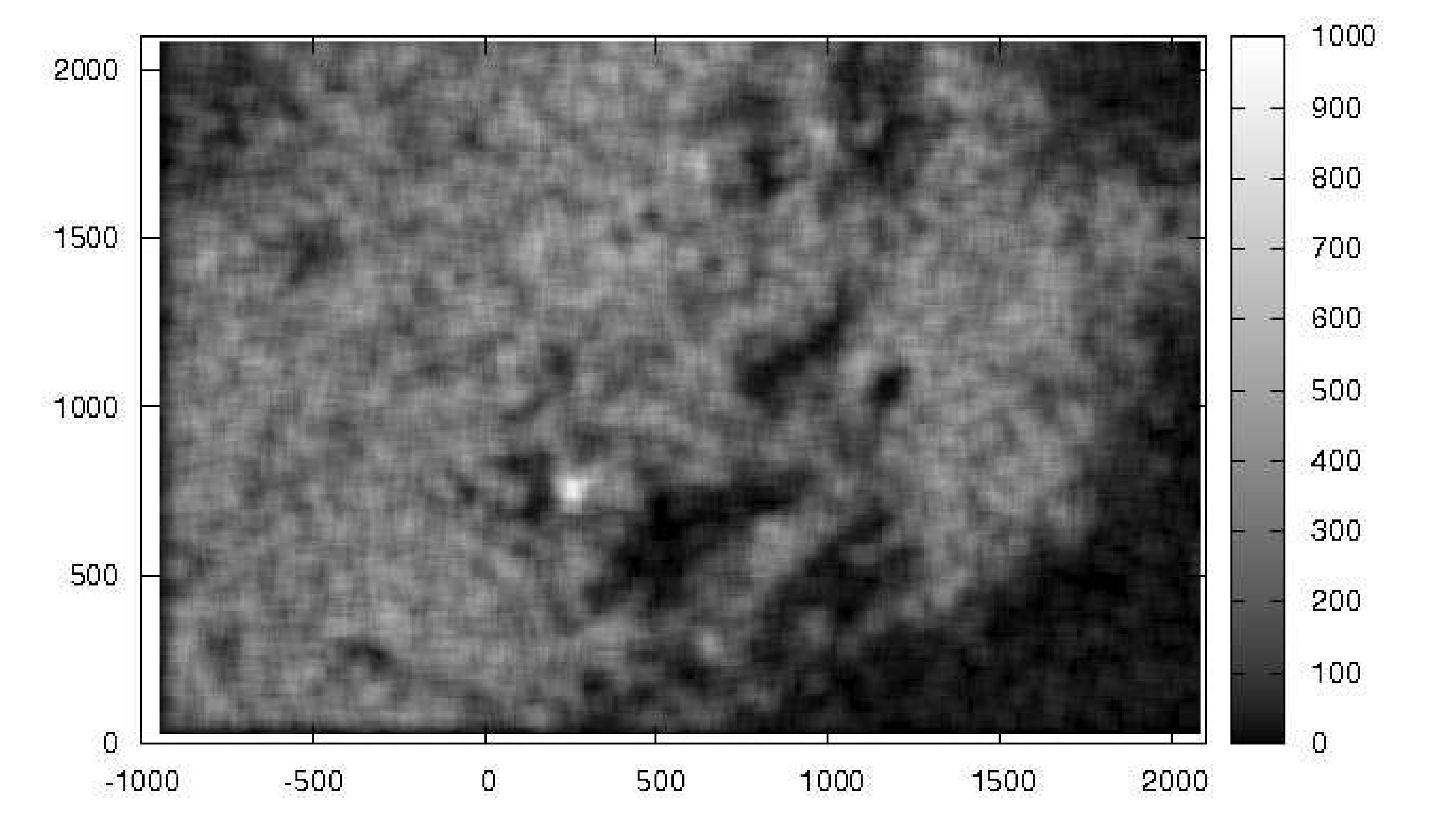}
\includegraphics[width=9cm]{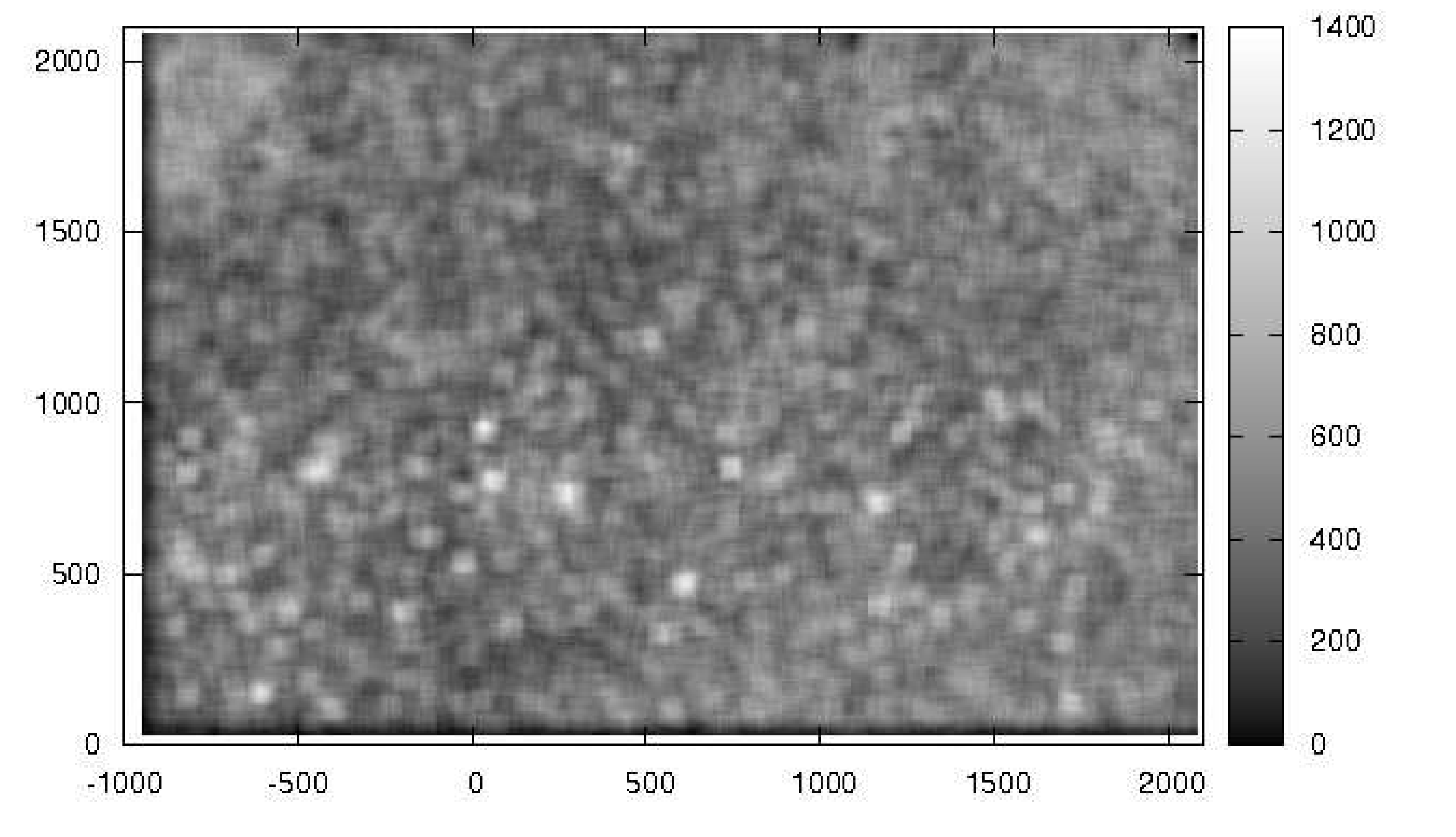}
\caption{{\it Upper panel}: Spatial density of stars with $K_\mathrm{s}\leq17$ located within
$\vert\Delta (J-K_\mathrm{s})\vert\leq0.07$ from the isochrone of Figure~\ref{f_isofit} for VVV~CL002.
The coordinates are given in pixels, and the density is in units of stars per arcmin$^{2}$. {\it Lower
panel}: same as the upper panel, but with an isochrone shifted 0.14 magnitudes toward the red.}
\label{f_denmapD2}
\end{center}
\end{figure}

The significance of the stellar overdensity detected in the direction of VVV~CL002 is the highest
among the three candidates. As an additional proof that the cluster and field stars originate from
different populations, we derived the spatial density distribution of the stars with $K_\mathrm{s}\leq$17
located within $\vert\Delta (J-K_\mathrm{s})\vert\leq$0.07 from the isochrone plotted in
Figure~\ref{f_isofit}, which fits the decontaminated cluster sequence with the parameters derived in
\S\ref{c_paramD2}. The resulting density map is shown in the upper panel of Figure~\ref{f_denmapD2},
where each 2$\arcsec \times 2\arcsec$ pixel is assigned the average density in the
20$\arcsec \times 20\arcsec$ square centered on it. The local overedensity associated to the cluster
is clearly visible at (x,y)$\approx$(250,750). The procedure was then repeated, shifting the same
isochrone by 0.14 magnitudes toward the red. The map is shown in the lower panel of the same figure,
where the cluster overdensity is barely visible, lost in the patchy structures of the field. A
random overdensity of stars should be observed in any region of the CMD where the field is found though.
This demonstrates that VVV~CL002 is not a simple fluctuation of the field density. Still, we
could be in the presence of a narrow, deep window in the interstellar clouds, where a higher quantity of
stars, on average bluer because less reddened, would be found. However, the presence of a similar color
offset even in the reddening-free CMD argues against this hypothesis. Therefore, VVV~CL002 and VVV~CL003
are most likely two new stellar clusters.

Even VVV~CL004 is a simple case, because all evidence points at a random fluctuation of field
stars. Indeed, local stellar clustering is not very significant, its stars are photometrically
indistinguishable from the field population, and its radial density profile, although well fitted with a
King profile, is quite peculiar. Nevertheless, VVV~CL004 could still be an open cluster remnant (OCR), a
dynamically evolved cluster in the last stage before dissolution. Open cluster remnants should appear as
small groups of bright stars, depleted of the fainter main\--sequence objects
\citep{delaFuente97,delaFuente98}, as can be observed in the field\--decontaminated CMD of VVV~CL004.

In the following section we will estimate the physical parameters of VVV~CL002 and VVV~CL003. The
results are summarized in Table~\ref{t_res}. A similar analysis is not presented for VVV~CL004, both because
the object is most probably not a real cluster, and because the stars in the decontaminated CMD are too few
and dispersed for a reliable estimate. The most satisfactory isochrone fit of the VVV~CL004 CMD is shown in
Figure~\ref{f_isofit}. It was obtained assuming $[\mathrm{Fe/H}]=-$1.1, $[\mathrm{\alpha /Fe}]=+0.30$,
(m$-$M)$_{K_\mathrm{s}}$=15.4, E($J-K_\mathrm{s}$)=0.82, and an age of 7.5 Gyr. It is not the only solution,
anyway, and other similar fits can be obtained with a very different set of parameters.

The transformations of \citet{Grocholski02} were used to link the theoretical $K$ values to the observed
$K_\mathrm{s}$ magnitudes. We relied on key photometric quantities measured on the decontaminated
CMDs, through the calibrations of \citet{Valenti04} and \citet[][hereafter F06]{Ferraro06}.
\defcitealias{Ferraro06}{F06} In particular, \citetalias{Ferraro06} define two sets of equations
for disk/halo-like and bulge-like objects, depending on the assumed trend of the $\alpha$-enhancement
with metallicity. Indeed, the $\alpha$-element abundance affects the location and curvature of the
RGB, which are better described as a function of the so-called global metallicity
($[\mathrm{M} /\mathrm{H}]$) which, for $[\mathrm{Fe} /\mathrm{H}]\leq -1$, can be expressed as
\begin{equation}
[\mathrm{M}/ \mathrm{H}]=[\mathrm{Fe} /\mathrm{H}]+\log{(0.638 f_\alpha +0.362)},
\label{e_MH}
\end{equation}
where $f_\alpha$ is the $\alpha$-elements enhancement factor \citep{Salaris93}. This relation,
however, breaks down with increasing metallicity \citep{VandenBerg00,Kim02}.

\subsection{VVV~CL002 parameters}
\label{c_paramD2}

The decontaminated CMD of VVV~CL002 shows a clump of stars at $K_\mathrm{s}\approx$14. We
interpret this as a signature of the cluster's horizontal branch red clump. Following the prescriptions
of \citet{Kuchinski95} and \citet{Kuchinski95b}, we linearly fitted the upper RGB from 0.6 mag above
the clump up to the brightest stars, obtaining a slope of $-$0.110. The fit
is shown in Figure~\ref{f_cmds}. This value is very similar to the slope of the
RGB of NGC\,6528 \citep{Ferraro00} and Liller\,1 \citep{Valenti10}, two metal-rich bulge globular
clusters enriched in $\alpha$-elements ($[\mathrm{\alpha /Fe}]\approx+0.30$). NGC\,6528 is the bulge
GC richest in iron, with $[\mathrm{Fe /H}]$ estimates in the range $-$0.1--\,+0.1
\citep{Zoccali04,Origlia05,Carretta09}, while \citet{Origlia01} report $[\mathrm{Fe/H}]=-0.30$ for
Liller\,1. For VVV~CL002 we used the \citetalias{Ferraro06} equations for bulge systems, because of
its location very close to the Galactic center (see later), obtaining $[\mathrm{M/H}]=-0.16\pm0.20$
and, assuming $[\mathrm{\alpha /Fe}]=+0.30$, $[\mathrm{Fe/H}]=-0.4\pm0.2$ from Equation~(\ref{e_MH}).
The errors were derived from the uncertainties in the equation constants estimated by \citet{Ferraro00},
which dominate the error budget when compared to the uncertainties in the derived slope. Had we used
the equations for disk/halo objects, we would have obtained $[\mathrm{M/H}]=-0.24$ and
$[\mathrm{Fe/H}]=-0.45$, consistent with the adopted values.

The mean magnitude and color of the stars in the red clump of VVV~CL002 are $K_\mathrm{s}=13.80\pm0.15$
and ($J-K_\mathrm{s})=2.17\pm0.04$, where the rms of the measurements are taken as uncertainties on the
average values. The absolute magnitude of the red clump weakly depends on the age and metallicity of the
stellar population. \citet{Ferraro06} propose M$_{K,0}$=$-1.40$ and a conservative error to take
this effect into account, but still their calibration is more suited for intermediate metal-poor, old
clusters. VVV~CL002 is at the upper limit of the metallicity range considered by \citetalias{Ferraro06},
and its red clump is therefore expected to be brighter,
as indicated by the theoretical values tabulated by \citet{Salaris02}. Indeed, \citet{Alves00} report
a mean value 0.2 magnitudes brighter than \citetalias{Ferraro06} for their solar-metallicity field
sample. We finally adopted M$_{K,0}$=$-1.50\pm0.15$ which, as shown by Table~1 of \citet{Salaris02},
encompasses the theoretical expectations for stars at $[\mathrm{Fe/H}]=-0.4$ in the age interval from
1 to 11 Gyr. We thus obtain the apparent distance modulus (m$-$M)$_{K_\mathrm{s}}=15.34\pm0.21$. From
Equation~3.4 of \citetalias{Ferraro06}, it follows that the apparent magnitude at the unreddened
color ($J-K_\mathrm{s})_0$=0.7 is $K_\mathrm{s}$=13.52. The observed color at this magnitude was
obtained with a linear fit of the cluster RGB, restricted to the range $K_\mathrm{s}$=13-14 to limit
the effects of the RGB curvature. The comparison of this value with the expectation for null reddening
gave E($J-K_\mathrm{s})=1.50\pm$0.15. The uncertainty associated to the fit returned a too small
error ($\pm$0.03 magnitudes), while the reddening of Galactic GCs is usually uncertain to at least 10\%
\citep[see, for example, the discussion in][2010 version]{Harris96}. The error on E($J-K_\mathrm{s}$) was
therefore arbitrarily fixed to the more realistic value given above. The reddening agrees well
with the value obtained by horizontally shifting a PGPUC isochrone (\citealt{Valcarce11};
Valcarce et al. 2011, {\it in preparation}), with the required metallicity, $\alpha$-enhancement, and
apparent distance modulus, to find the best fit of the observed RGB. Through the transformations of
\citet{Cardelli89}, the derived reddening translates into E($B-V$)=2.88$\pm$0.29,
A$_{K_\mathrm{s}}=1.02\pm$0.10, implying a true distance modulus (m$-$M)$_0$=14.32$\pm$0.23 and a
distance d=7.3$\pm$0.9~kpc. Had we used the mean intrinsic color of the red clump proposed by
\citet{Lopez02}, we would have obtained E($J-K_\mathrm{s})=1.46\pm$0.15, in good agreement with our
estimate, but in this case the final isochrone fit of the cluster CMD is of lesser quality.
\citet{Schlegel98}'s maps give E($B-V$)=3.47 in the direction of VVV~CL002, but this higher value should
not be surprising, because it refers to the total Galactic absorption, while the cluster should be located
only half way to the end of the absorbing gas column. In any case, the corrections proposed by
\citet{Bonifacio00} give E($B-V$)=2.29, a value incompatible with our results.

At a distance of 7.3 kpc, the tidal and half-light angular radii of VVV~CL002 (\S\ref{c_radprof})
translate into the physical dimensions 3.8$\pm$0.5 and 1.6$\pm$0.3 pc, respectively. From the estimated
distance and the Galactic coordinates ($l,b$), assuming a solar Galactocentric distance R$_\odot$=8~kpc,
we also derived the distance from the Galactic center R$_\mathrm{GC}$=0.6$\pm$0.8~kpc, and the height
above the Galactic plane $z$=113$\pm$12~pc.

The cluster age was determined by fitting the lower part of the decontaminated CMD with a PGPUC
isochrone with the metallicity, $\alpha$-enhancement, distance modulus, and reddening determined above.
In Figure~\ref{f_isofit} we show this fit with two isochrones of different age. We found that a young
age ($\sim$6.5 Gyr) was required to fit the feature at
($K_\mathrm{s},(J-K_\mathrm{s}))\approx (17,2)$. This clump of stars, closely resembling the cluster's
sub-giant branch (SGB) and upper turn-off, was always observed when varying the cluster area and the
comparison field in the decontamination procedure. However, we regard this age estimate as uncertain,
because it completely relies on stars at the fainter end of our diagram, but it can be regarded as a
lower limit: had the cluster been younger, the SGB would have been brighter, and thus more clearly
observed in our photometry. Deeper data unveiling the turnoff region are required to confirm these
results. It must also be remembered that we used isochrones with Y=0.245, but a different helium
abundance can affect the age estimate, because stars with higher helium content evolve faster.

The cluster distance and all other distance-dependent parameters such as R$_\mathrm{GC}$, $z$,
and the physical size, depend on the assumed absolute magnitude of the red clump. As an independent
test of our results, we estimated the distance modulus differentially with respect to 47\,Tucanae
($[\mathrm{Fe/H}]=-0.7$, \citealt{Gratton03}), and NGC\,6528 ($[\mathrm{Fe/H}]\approx 0$), assuming the
photometry by \citet{Valenti04b} to derive the shift in color and magnitude between them and VVV~CL002,
and the \citet{Cardelli89} extinction coefficients. Despite the different metallicity of the two
comparison clusters, the resulting distance modulus differs by only one tenth of magnitude in the two
cases. We obtain (m$-$M)$_0$=14.16, and a distance d=6.8~kpc. that agrees within errors with our
estimate. The difference is caused by the old age of the two comparison clusters
\citep[10--11~Gyr,][]{Gratton03,Ortolani01}: the distance modulus derived with this empirical method
is 0.16 magnitudes lower than what we previously obtained, and \citet{Salaris02} predict that the red
clump of a population of similar age and $[\mathrm{Fe/H}]=-0.4$ would be 0.12--0.15 magnitudes fainter
than our assumption. We thus conclude that, although the two estimates agree well, a distance shorter by
about 0.5~kpc should be preferred if VVV~CL002 is an old GC.

We estimated the total luminosity of the cluster integrating the flux collected within increasing
apertures up to the tidal radius, and subtracting the expected field contribution. The measurements were
performed with SExtractor \citep{Bertin96} on the stacked tile in the $J$\--band, because many saturated
stars prevented a reliable estimate in the $K_\mathrm{s}$-band frame. We thus found a total magnitude
$J_\mathrm{tot}=12.5\pm$0.2, and from the mean cluster color ($J-K_\mathrm{2})=2.3\pm$0.1 and the
apparent distance modulus, we obtain M$_Ks=-5.1\pm$0.3, which translates into M$_V=-3.4\pm$0.3
assuming the synthetic colors of an old stellar population calculated by \citet{Leitherer99}.

\subsection{VVV~CL003 parameters}
\label{c_paramD3}

The interpretation of the decontaminated CMD of VVV~CL003 must take into account the problems that its
photometry presented. The diagram reveals a clear sequence of
stars distinct from the field population, resembling a cluster RGB. A sparse group of highly reddened
stars is also visible to the right of this sequence. Many of them overlap with the cluster RGB in
the reddening-free CMD and are probably the result of differential reddening, while some are most
probably foreground cool dwarfs, because they still keep their redder color even in Figure~\ref{f_marcindex}.
On the bluer side, a residual contamination is found. There is a lack of stars fainter than
$K_\mathrm{s}$=16, probably because almost the entire cluster area falls in the CCD border,
where the data were collected in only one of the two jittered exposures forming a frame. The cluster
RGB and the redder objects are visible for any definition of cluster and field area, provided that the
latter is not more distant than 2-3$\arcmin$ from the center, while the residual contamination may
change in color and quantity of stars depending on the location of the comparison field.

\begin{figure}
\begin{center}
\includegraphics[width=9cm]{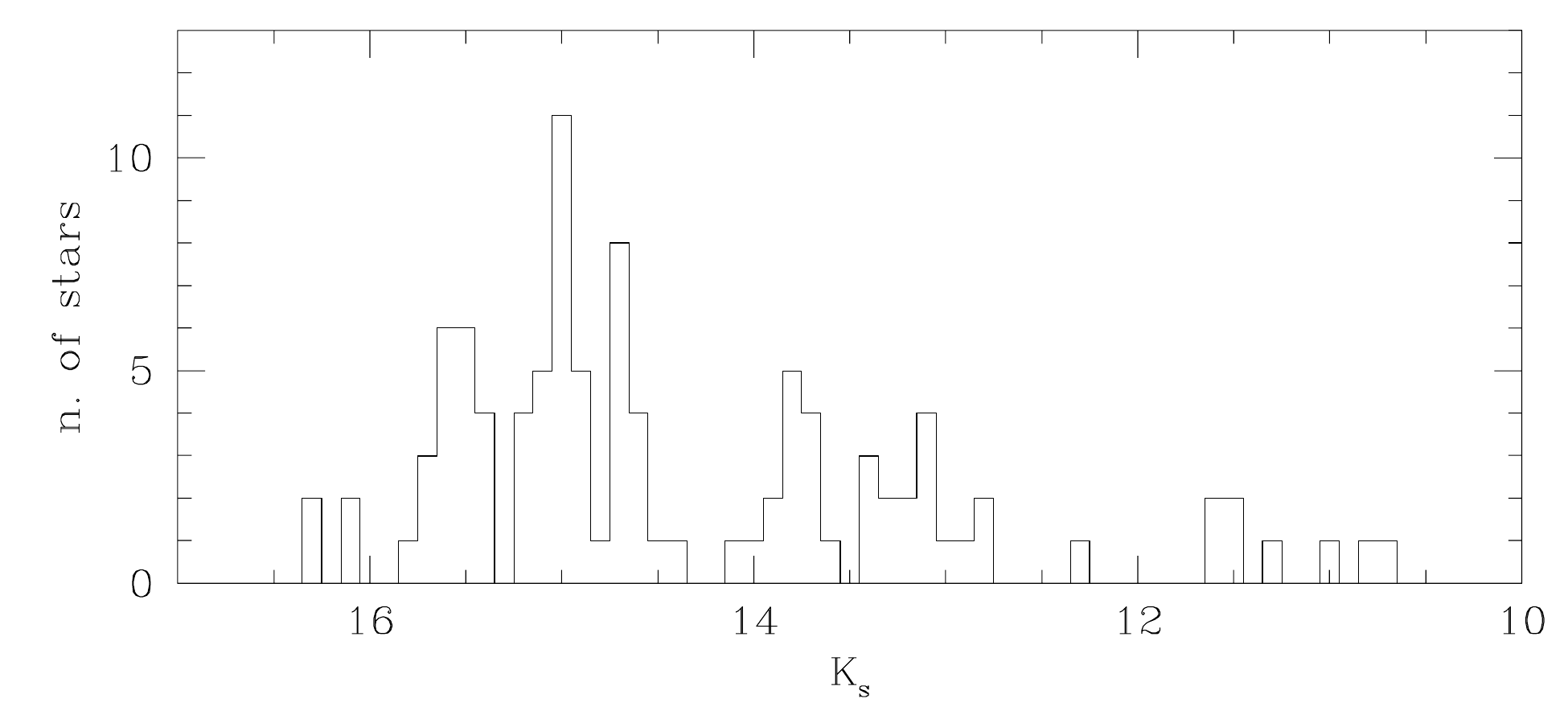}
\caption{Luminosity function of RGB stars in VVV~CL003.}
\label{f_rcD3}
\end{center}
\end{figure}

\begin{figure}
\begin{center}
\includegraphics[width=9cm]{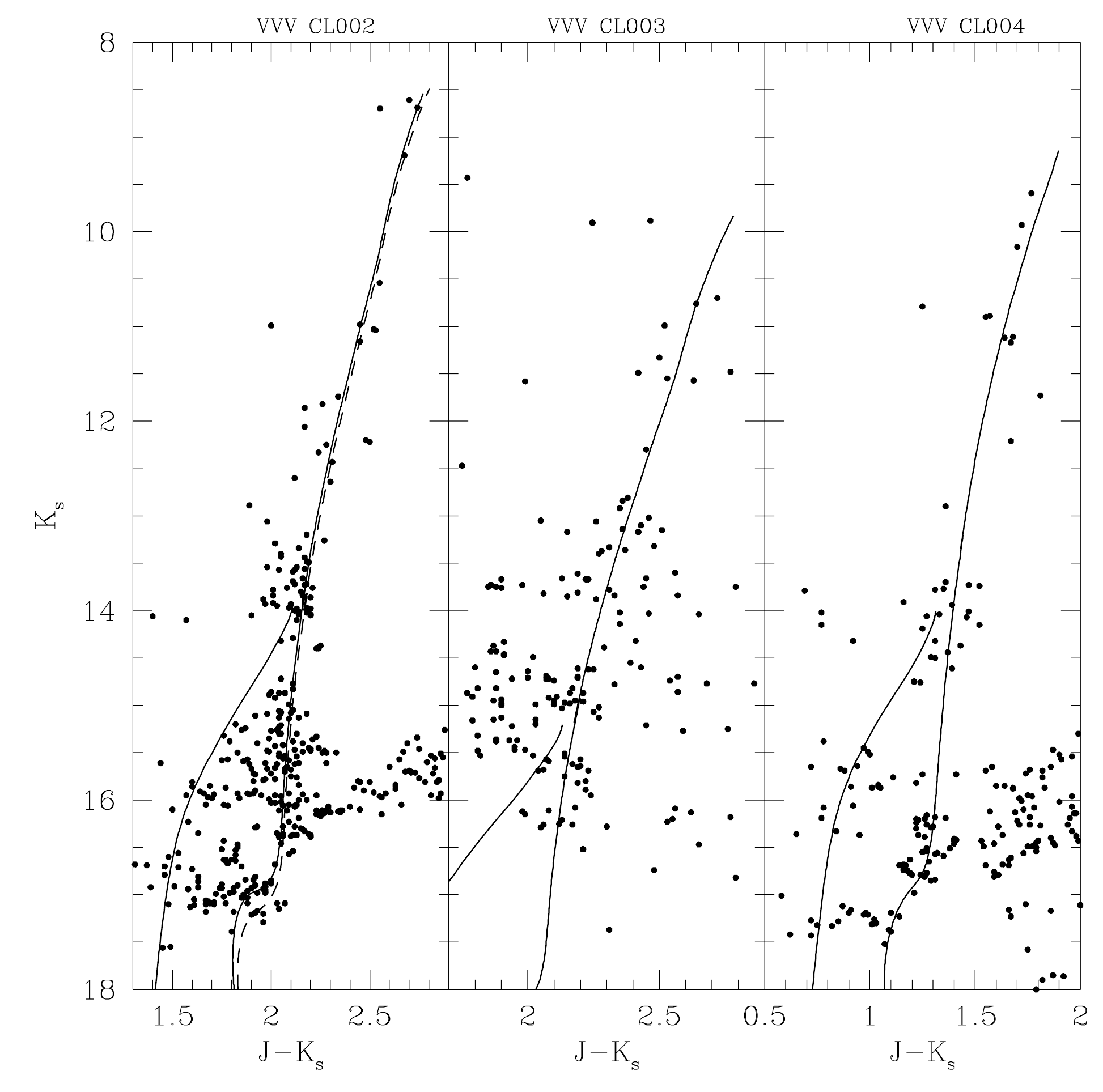}
\caption{Decontaminated CMDs of the three clusters, with the best-fitting PGPUC isochrones defined in
the text overplotted. The zero-age horizontal branch (ZAHB) in all panels of the figure
was calculated with the helium abundance Y=0.245. For VVV~CL002, two isochrones at 6.5~Gyr (thick curve)
and 8.5~Gyr (dashed curve) are shown.}
\label{f_isofit}
\end{center}
\end{figure}

A dense group of stars in the cluster RGB is found at $K_\mathrm{s}$=15. This feature is more
easily observed plotting the histogram of the distribution of stars along the sequence, as shown in
Figure~\ref{f_rcD3}. In this plot we considered only stars within 0.2 magnitudes in color from the RGB
ridge line, defined as the isochrone plotted in Figure~\ref{f_isofit}. Unfortunately, the field and
cluster RGBs are close in this part of the CMD, and the behavior of this feature (and even its
detection) strongly depends on the choice of the comparison field. However, an enhanced residual
contamination appeared at the same magnitude and a bluer color when the clump in the RGB was not found,
indicating that the field RGB was too red, partially overlapping and erasing
the cluster clump. We will therefore assume this group of stars is the cluster horizontal branch red
clump, but we caution that the subsequent analysis is based on this detection, which is not clear beyond
any doubt.

Repeating the same procedures used for VVV~CL002 (\S\ref{c_paramD2}), the slope of the upper RGB
estimated with the fit shown in the right panel of Figure~\ref{f_cmds} is $-$0.117. This value is
lower than the slope of any of the 27 Galactic GCs studied by \citet{Ferraro00} and
\citet{Valenti04,Valenti10},
indicating a high metal content. The location of VVV~CL003 in the Galactic disk outside the bulge
(see below) led us to adopt the \citetalias{Ferraro06} calibration for disk objects and no
$\alpha$-enhancement, obtaining $[\mathrm{M/H}]=[\mathrm{Fe/H}]=-0.1\pm0.2$. For the absolute
magnitude of the red clump we opted for the  value proposed by \citet{Alves00}, given the high
metallicity derived from the RGB slope. We thus obtained an apparent distance modulus
(m$-$M)$_{K_\mathrm{s}}$=16.58$\pm$0.12, E($J-K_\mathrm{s})=1.48\pm$0.15 from Equation~2.4 of
\citetalias{Ferraro06}, and E($B-V$)=2.85$\pm$0.29. Using the calibration of \citet{Lopez02}, we
would have obtained E($J-K_\mathrm{s}$)=1.46. On the contrary, \citet{Schlegel98} give E($B-V$)=3.34
in the direction of VVV~CL003. As in the case of VVV~CL002, this value is too high compared with our
estimates, while the corrections of \citet{Bonifacio00} reduce it too much, returning E($B-V$)=2.19.
However, the resolution of the \citet{Schlegel98} maps is too low for this region of the sky of strong
reddening gradients, and it is not surprising that their average value in the resolution element does
not coincide with the local absorption. It could also be that the relation
R=A$_\mathrm{V}$/E($B-V$)=3.1, required to transform the reddening in different bands through the
relations of \citet{Cardelli89}, is not valid in this highly obscured part of the sky.

The reddening in the infrared bands implies A$_{K_\mathrm{s}}=1.01\pm$0.10, a true distance modulus
(m$-$M)$_0$=15.57$\pm$0.16, and a distance d=13$\pm$1.0 kpc. With simple geometrical arguments
based on the ($l$,$b$) coordinates given in Table~\ref{t_coord}, we obtain the Galactic height
$z$=166$\pm$14~pc, and the Galactocentric distance R$_\mathrm{GC}$=5.0$\pm$1.0~kpc. At a
heliocentric distance of 13~kpc, the angular size of VVV~CL003 translates into the physical quantities
r$_\mathrm{t}=6.8\pm$0.4 pc and r$_\mathrm{h}=2.3\pm$0.4 pc.

\begin{table}[t]
\begin{center}
\caption{Derived parameters of the two cluster candidates.}
\label{t_res}
\begin{tabular}{l | c c}
\hline
\hline
 & VVV~CL002 & VVV~CL003 \\
\hline
r$_\mathrm{h}$ ($\arcmin$) & 0.75$\pm$0.10  & 0.6$\pm$0.1 \\
r$_\mathrm{h}$ (pc)        & 1.6$\pm$0.3    & 2.3$\pm$0.4 \\
r$_\mathrm{t}$ ($\arcmin$) & 1.8$\pm$0.1    & 1.8$\pm$0.1 \\
r$_\mathrm{t}$ (pc)        & 3.8$\pm$0.5    & 6.8$\pm$0.4 \\
$c=\log{(r_\mathrm{h}/r_\mathrm{c})}$ & 0.65$\pm$0.26 & 0.56$\pm$0.21 \\
(m$-$M)$_0$                & 14.32$\pm0.23$ & 15.57$\pm$0.16 \\
d (kpc)                    & 7.3$\pm$0.9    & 13.0$\pm$1.0 \\
R$_\mathrm{GC}$ (kpc)      & 0.7$\pm$0.9    & 5.0$\pm$1.0 \\
$z$ (pc)                   & 113$\pm$13     & 166$\pm$14 \\
E($J-K_\mathrm{s}$)        & 1.50$\pm$0.15  & 1.48$\pm$0.15 \\
E($B-V$)                   & 2.88$\pm$0.29  & $2.85\pm0.29$ \\
$[\mathrm{M/H}]$           & $-0.16\pm0.2$  & $-0.1\pm0.2$ \\
$[\mathrm{Fe/H}]$          & $-0.4\pm0.2$   & $-0.1\pm0.2$ \\
M$_V$                      & $-$3.4$\pm$0.3 & - \\
Age (Gyr) & $\geq$6.5 & - \\
\hline
\end{tabular}
\end{center}
\end{table}

An age estimate for VVV~CL003 is impossible, because the CMD does not reach the SGB. We
found that the observed RGB could be satisfactorily fitted with isochrones of virtually any age,
although lower values required a progressively higher reddening to match the observed sequence. For
example, an isochrone of 2 Gyr better fitted the cluster CMD if E($J-K_\mathrm{s}$)=1.56 is assumed.
This is not necessarily a reason to prefer an older age for VVV~CL003, because it could simply reflect
the fact that the \citetalias{Ferraro06} relation used to estimate the reddening was calibrated on old
stellar populations.

The high interstellar reddening gradient and the problematic photometric data (\S\ref{c_reduction})
prevented a reliable estimate of the cluster total luminosity.


\section{Discussion and conclusions}
\label{c_conclusions}

The age of VVV~CL002 is uncertain, but the proposed lower limit of 6.5 Gyr, its subsolar metallicity,
and its height above the Galactic plane, twice the scale height of young objects \citep{Joshi05}, argue in
favor of an old cluster. On the other hand, \citet{Friel95} showed that old open clusters are completely
absent in the inner $\sim$7~kpc from the Galactic center. Because of its location very close to the
Galactic center, VVV~CL002 is therefore most likely a GC.

\begin{figure*}
\begin{center}
\includegraphics[width=18cm]{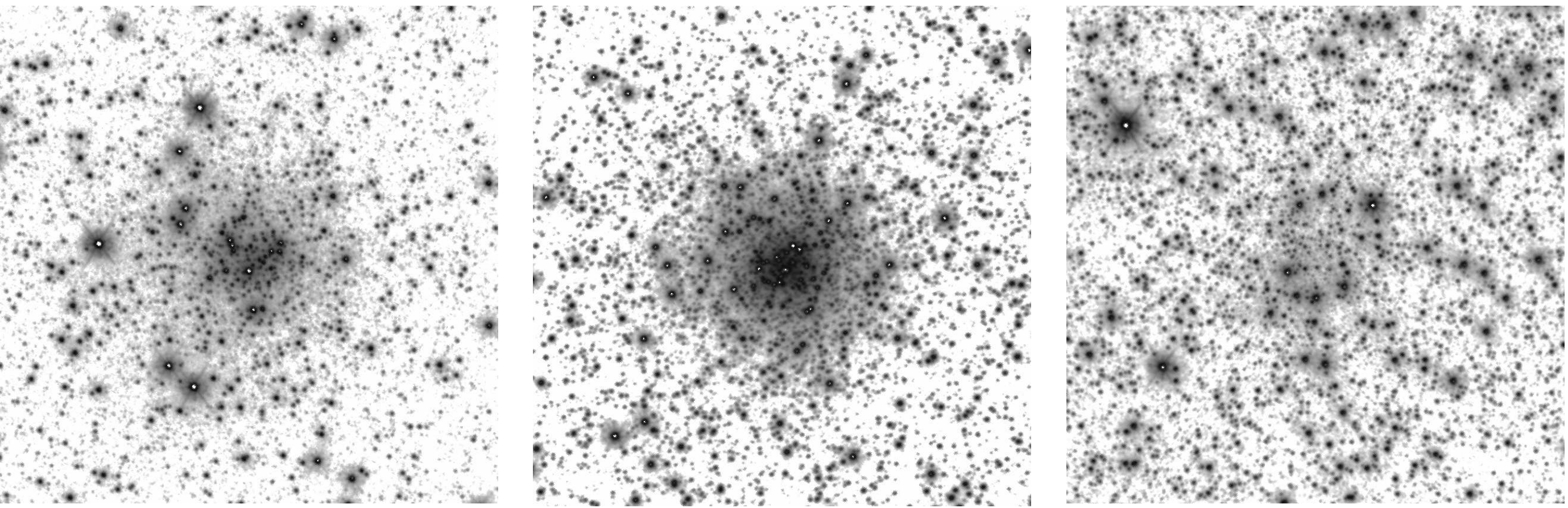}
\caption{$2\farcm 4 \times 2\farcm 4$ logarithmic gray-scale $K_\mathrm{s}$-band images of (from left to
right) NGC\,6528, Liller\,1, and VVV~CL002, from the VVV archive. North is up and East is to the right.}
\label{f_compGCs}
\end{center}
\end{figure*}

So far, HP\,1 is the closest known GC to the Galactic center \citep[R$_\mathrm{GC}=$0.5~kpc][]{Barbuy06},
and only six GCs have been discovered in the inner 1~kpc \citep{Harris96}. VVV~CL002, with an estimated
Galactocentric distance of 0.7~kpc, is therefore one of the innermost Galactic GCs. Curiously, the
density profile of VVV~CL002 suggests the presence of tidal tails, a behavior more easily observed
when the cluster is close to the perigalacticon \citep{Kuepper10b}. If the connection between tidal
tails and perigalacticon were to be confirmed even for objects subject to the intense stress of the inner
Galaxy, then the cluster orbit would be confined to the very central region of the Galaxy.

VVV~CL002 has a global metallicity close to NGC\,6528 and Liller\,1, as indicated by the slope of the
upper RGB. These two objects are also bulge GCs very close to the Galactic center
\citep[R$_\mathrm{GC}$=0.6 and 0.8 kpc, respectively;][]{Valenti10,Feltzing02} and, as in the case of
VVV~CL002, their stars also are barely distinguished from the background bulge population in the CMD
\citep{Davidge00}. However, the similarity cannot be drawn further, because their appearance is very
distinct. In Figure~\ref{f_compGCs} we show a visual comparison between the three objects, with
$K_\mathrm{s}$-band images from the VSA archive: VVV~CL002 is clearly sparser and less massive,
resembling a Palomar-like object more than a highly concentrated bulge GC such as NGC\,6528 and
Liller\,1. Indeed, our estimate of its total luminosity is $\sim$3.5 magnitudes fainter than these
two objects \citep{Webbink85,Peterson87,vandenBerg91,Mallen93}. VVV~CL002 is more affine to moderately
metal-rich, low-mass, and sparse GCs such as Whiting\,1 and Palomar\,1, whose physical size are comparable
\citep[r$_\mathrm{h}$=1.9 and 1.5-2.2 pc, respectively;][]{Harris96,Rosenberg98}. These two objects
are relatively young \citep{Rosenberg98,Carraro05}, and are only slightly more metal-poor
\citep{Carraro07a,Monaco11}, although they are one magnitude brighter \citep{Harris96}. Palomar\,12
also has a similar young age \citep{Stetson89,Geisler07} but has a lower metallicity \citep{Cohen04}, and
it is physically larger \citep{Harris96}. It is worth noting that both Whiting\,1 and Palomar\,12 are
associated with the Sagittarius dwarf spheroidal galaxy \citep{Irwin99,Carraro07a,Law10}, while the
extragalactic origin of Palomar\,1 has been repeatedly proposed although never proven
\citep[e.g.,][]{Crane03,Frinchaboy04,Belokurov07}.

VVV~CL003 is surely a physical cluster of stars, located in the Galactic disk beyond the center,
outside the bulge, at a Galactocentric distance of 5~kpc. Its high metallicity and its position in the
Galaxy are more typical of an open cluster, although a GC cannot be excluded. Deeper photometry,
reaching at least the cluster SGB two magnitudes fainter than the limit reached with our data, is
needed to estimate its age and, thus, better clarify its nature. Interestingly enough, to our knowledge
it is the first stellar cluster ever discovered on the ``dark side of the Galaxy", i.e. the Galactic
disk on the other side of the center. The Open Cluster Database\footnote{http://www.univie.ac.at/webda/}
\citep[WEBDA,][]{Mermilliod96} has no entry in the direction of the bulge more distant than a few kpc.

VVV~CL004 is most probably only an asterism, and as such it deserves no further attention. Nevertheless,
the possibility of an OCR remains open, and this kind of object is of great interest, because we are still
lacking observational evidence of the last stages of the cluster dynamical evolution. Indeed, OCRs are
very elusive and easily confused with random field fluctuations and, despite a large list of candidates
\citep[e.g.,][]{Pavani07,Bica01}, only one genuine OCR has been to date spectroscopically confirmed
\citep[NGC\,1901,][]{Carraro07}.


\begin{acknowledgements}

We gratefully acknowledge use of data from the ESO Public Survey programme ID~179.B-2002 taken with the
VISTA telescope, and data products from the Cambridge Astronomical Survey Unit. We acknowledge support by
the FONDAP Center for Astrophysics 15010003, BASAL Center for Astrophysics and Associated Technologies
PFB-06/2007, the Chilean Ministry for the Economy, Development, and Tourism's Programa Iniciativa
Cient\'{i}fica Milenio through grant P07-021-F, awarded to The Milky Way Millennium Nucleus, FONDECYT 1090213,
1080086, 1110393 and 1110326, and from CONICYT. This investigation made use of data from the Two Micron
All Sky Survey, which is a joint project of the University of Massachusetts and the Infrared Processing and
Analysis Center/California Institute of Technology, funded by the National Aeronautics and Space
Administration and the National Science Foundation. We thank the referee, W.~E. Harris, for improving
the manuscript quality with his detailed comments.
\end{acknowledgements}


\bibliographystyle{aa}
\bibliography{D2bib.bib}

\begin{thebibliography}{92}
\expandafter\ifx\csname natexlab\endcsname\relax\def\natexlab#1{#1}\fi

\bibitem[{{Abazajian} {et~al.}(2009){Abazajian}, {Adelman-McCarthy},
  {Ag{\"u}eros}, {Allam}, {Allende Prieto}, {An}, {Anderson}, {Anderson},
  {Annis}, {Bahcall}, \& et~al.}]{Abazajian09}
{Abazajian}, K.~N., {Adelman-McCarthy}, J.~K., {Ag{\"u}eros}, M.~A., {et~al.}
  2009, \apjs, 182, 543

\bibitem[{{Alves}(2000)}]{Alves00}
{Alves}, D.~R. 2000, \apj, 539, 732

\bibitem[{{Barbuy} {et~al.}(2006){Barbuy}, {Zoccali}, {Ortolani}, {Momany},
  {Minniti}, {Hill}, {Renzini}, {Rich}, {Bica}, {Pasquini}, \&
  {Yadav}}]{Barbuy06}
{Barbuy}, B., {Zoccali}, M., {Ortolani}, S., {et~al.} 2006, \aap, 449, 349

\bibitem[{{Belokurov} {et~al.}(2007){Belokurov}, {Evans}, {Irwin},
  {Lynden-Bell}, {Yanny}, {Vidrih}, {Gilmore}, {Seabroke}, {Zucker},
  {Wilkinson}, {Hewett}, {Bramich}, {Fellhauer}, {Newberg}, {Wyse}, {Beers},
  {Bell}, {Barentine}, {Brinkmann}, {Cole}, {Pan}, \& {York}}]{Belokurov07}
{Belokurov}, V., {Evans}, N.~W., {Irwin}, M.~J., {et~al.} 2007, \apj, 658, 337

\bibitem[{{Benjamin} {et~al.}(2003){Benjamin}, {Churchwell}, {Babler}, {Bania},
  {Clemens}, {Cohen}, {Dickey}, {Indebetouw}, {Jackson}, {Kobulnicky},
  {Lazarian}, {Marston}, {Mathis}, {Meade}, {Seager}, {Stolovy}, {Watson},
  {Whitney}, {Wolff}, \& {Wolfire}}]{Benjamin03}
{Benjamin}, R.~A., {Churchwell}, E., {Babler}, B.~L., {et~al.} 2003, \pasp,
  115, 953

\bibitem[{{Bertin} \& {Arnouts}(1996)}]{Bertin96}
{Bertin}, E., \& {Arnouts}, S. 1996, \aaps, 117, 393

\bibitem[{{Bica} {et~al.}(2007){Bica}, {Bonatto}, {Ortolani}, \&
  {Barbuy}}]{Bica07}
{Bica}, E., {Bonatto}, C., {Ortolani}, S., \& {Barbuy}, B. 2007, \aap, 472, 483

\bibitem[{{Bica} {et~al.}(2001){Bica}, {Santiago}, {Dutra}, {Dottori}, {de
  Oliveira}, \& {Pavani}}]{Bica01}
{Bica}, E., {Santiago}, B.~X., {Dutra}, C.~M., {et~al.} 2001, \aap, 366, 827

\bibitem[{{Bonatto} {et~al.}(2007){Bonatto}, {Bica}, {Ortolani}, \&
  {Barbuy}}]{Bonatto07}
{Bonatto}, C., {Bica}, E., {Ortolani}, S., \& {Barbuy}, B. 2007, \mnras, 381,
  L45

\bibitem[{{Bonifacio} {et~al.}(2000){Bonifacio}, {Monai}, \&
  {Beers}}]{Bonifacio00}
{Bonifacio}, P., {Monai}, S., \& {Beers}, T.~C. 2000, \aj, 120, 2065

\bibitem[{{Borissova} {et~al.}(2011){Borissova}, {Bonatto}, {Kurtev}, {Clarke},
  {Pe{\~n}aloza}, {Sale}, {Minniti}, {Alonso-Garc{\'{\i}}a}, {Artigau},
  {Barb{\'a}}, {Bica}, {Baume}, {Catelan}, {Chen{\`e}}, {Dias}, {Folkes},
  {Froebrich}, {Geisler}, {de Grijs}, {Hanson}, {Hempel}, {Ivanov}, {Kumar},
  {Lucas}, {Mauro}, {Moni Bidin}, {Rejkuba}, {Saito}, {Tamura}, \&
  {Toledo}}]{Borissova11}
{Borissova}, J., {Bonatto}, C., {Kurtev}, R., {et~al.} 2011, \aap, 532, A131

\bibitem[{{Cardelli} {et~al.}(1989){Cardelli}, {Clayton}, \&
  {Mathis}}]{Cardelli89}
{Cardelli}, J.~A., {Clayton}, G.~C., \& {Mathis}, J.~S. 1989, \apj, 345, 245

\bibitem[{{Carraro}(2005)}]{Carraro05}
{Carraro}, G. 2005, \apjl, 621, L61

\bibitem[{{Carraro} {et~al.}(2007{\natexlab{a}}){Carraro}, {de La Fuente
  Marcos}, {Villanova}, {Moni Bidin}, {de La Fuente Marcos}, {Baumgardt}, \&
  {Solivella}}]{Carraro07}
{Carraro}, G., {de La Fuente Marcos}, R., {Villanova}, S., {et~al.}
  2007{\natexlab{a}}, \aap, 466, 931

\bibitem[{{Carraro} {et~al.}(2007{\natexlab{b}}){Carraro}, {Zinn}, \& {Moni
  Bidin}}]{Carraro07a}
{Carraro}, G., {Zinn}, R., \& {Moni Bidin}, C. 2007{\natexlab{b}}, \aap, 466,
  181

\bibitem[{{Carretta} {et~al.}(2009){Carretta}, {Bragaglia}, {Gratton},
  {D'Orazi}, \& {Lucatello}}]{Carretta09}
{Carretta}, E., {Bragaglia}, A., {Gratton}, R., {D'Orazi}, V., \& {Lucatello},
  S. 2009, \aap, 508, 695

\bibitem[{{Catelan} {et~al.}(2011){Catelan}, {Minniti}, {Lucas},
  {Alonso-Garcia}, {Angeloni}, {Beamin}, {Bonatto}, {Borissova}, {Contreras},
  {Cross}, {Dekany}, {Emerson}, {Eyheramendy}, {Geisler}, {Gonzalez-Solares},
  {Helminiak}, {Hempel}, {Irwin}, {Ivanov}, {Jordan}, {Kerins}, {Kurtev},
  {Mauro}, {Moni Bidin}, {Navarrete}, {Perez}, {Pichara}, {Read}, {Rejkuba},
  {Saito}, {Sale}, \& {Toledo}}]{Catelan11}
{Catelan}, M., {Minniti}, D., {Lucas}, P.~W., {et~al.} 2011, in Carnegie
Observatories Astrophysics Series, Vol. 5, RR Lyrae Stars, Metal-Poor Stars,
and the Galaxy, ed. A. McWilliam (arXiv:1105.1119)

\bibitem[{{Cohen}(2004)}]{Cohen04}
{Cohen}, J.~G. 2004, \aj, 127, 1545

\bibitem[{{Crane} {et~al.}(2003){Crane}, {Majewski}, {Rocha-Pinto},
  {Frinchaboy}, {Skrutskie}, \& {Law}}]{Crane03}
{Crane}, J.~D., {Majewski}, S.~R., {Rocha-Pinto}, H.~J., {et~al.} 2003, \apjl,
  594, L119

\bibitem[{{Davidge}(2000)}]{Davidge00}
{Davidge}, T.~J. 2000, \apjs, 126, 105

\bibitem[{{de La Fuente Marcos}(1997)}]{delaFuente97}
{de La Fuente Marcos}, R. 1997, \aap, 322, 764

\bibitem[{{de La Fuente Marcos}(1998)}]{delaFuente98}
{de La Fuente Marcos}, R. 1998, \aap, 333, L27

\bibitem[{{Elson} {et~al.}(1987){Elson}, {Fall}, \& {Freeman}}]{Elson87}
{Elson}, R.~A.~W., {Fall}, S.~M., \& {Freeman}, K.~C. 1987, \apj, 323, 54

\bibitem[{{Emerson} \& {Sutherland}(2010)}]{Emerson10}
{Emerson}, J., \& {Sutherland}, W. 2010, The Messenger, 139, 2

\bibitem[{{Feltzing} \& {Johnson}(2002)}]{Feltzing02}
{Feltzing}, S., \& {Johnson}, R.~A. 2002, \aap, 385, 67

\bibitem[{{Ferraro} {et~al.}(2000){Ferraro}, {Montegriffo}, {Origlia}, \& {Fusi
  Pecci}}]{Ferraro00}
{Ferraro}, F.~R., {Montegriffo}, P., {Origlia}, L., \& {Fusi Pecci}, F. 2000,
  \aj, 119, 1282

\bibitem[{{Ferraro} {et~al.}(2006){Ferraro}, {Valenti}, \&
  {Origlia}}]{Ferraro06}
{Ferraro}, F.~R., {Valenti}, E., \& {Origlia}, L. 2006, \apj, 649, 243

\bibitem[{{Friel}(1995)}]{Friel95}
{Friel}, E.~D. 1995, \araa, 33, 381

\bibitem[{{Frinchaboy} {et~al.}(2004){Frinchaboy}, {Majewski}, {Crane}, {Reid},
  {Rocha-Pinto}, {Phelps}, {Patterson}, \& {Mu{\~n}oz}}]{Frinchaboy04}
{Frinchaboy}, P.~M., {Majewski}, S.~R., {Crane}, J.~D., {et~al.} 2004, \apjl,
  602, L21

\bibitem[{{Froebrich} {et~al.}(2008{\natexlab{a}}){Froebrich}, {Meusinger}, \&
  {Davis}}]{Froebrich08}
{Froebrich}, D., {Meusinger}, H., \& {Davis}, C.~J. 2008{\natexlab{a}}, \mnras,
  383, L45

\bibitem[{{Froebrich} {et~al.}(2008{\natexlab{b}}){Froebrich}, {Meusinger}, \&
  {Scholz}}]{Froebrich08b}
{Froebrich}, D., {Meusinger}, H., \& {Scholz}, A. 2008{\natexlab{b}}, \mnras,
  390, 1598

\bibitem[{{Gallart} {et~al.}(2003){Gallart}, {Zoccali}, {Bertelli}, {Chiosi},
  {Demarque}, {Girardi}, {Nasi}, {Woo}, \& {Yi}}]{Gallart03}
{Gallart}, C., {Zoccali}, M., {Bertelli}, G., {et~al.} 2003, \aj, 125, 742

\bibitem[{{Geisler} {et~al.}(2007){Geisler}, {Wallerstein}, {Smith}, \&
  {Casetti-Dinescu}}]{Geisler07}
{Geisler}, D., {Wallerstein}, G., {Smith}, V.~V., \& {Casetti-Dinescu}, D.~I.
  2007, \pasp, 119, 939

\bibitem[{{Gnedin} \& {Ostriker}(1997)}]{Gnedin97}
{Gnedin}, O.~Y., \& {Ostriker}, J.~P. 1997, \apj, 474, 223

\bibitem[{{Gratton} {et~al.}(2003){Gratton}, {Bragaglia}, {Carretta},
  {Clementini}, {Desidera}, {Grundahl}, \& {Lucatello}}]{Gratton03}
{Gratton}, R.~G., {Bragaglia}, A., {Carretta}, E., {et~al.} 2003, \aap, 408,
  529

\bibitem[{{Grocholski} \& {Sarajedini}(2002)}]{Grocholski02}
{Grocholski}, A.~J., \& {Sarajedini}, A. 2002, \aj, 123, 1603

\bibitem[{{Harris}(1996)}]{Harris96}
{Harris}, W.~E. 1996, \aj, 112, 1487

\bibitem[{{Irwin}(1999)}]{Irwin99}
{Irwin}, M. 1999, in IAU Symposium, Vol. 192, The Stellar Content of Local
  Group Galaxies, ed. {P.~Whitelock \& R.~Cannon}, 409

\bibitem[{{Irwin} {et~al.}(2004){Irwin}, {Lewis}, {Hodgkin}, {Bunclark},
  {Evans}, {McMahon}, {Emerson}, {Stewart}, \& {Beard}}]{Irwin04}
{Irwin}, M.~J., {Lewis}, J., {Hodgkin}, S., {et~al.} 2004, in Society of
  Photo-Optical Instrumentation Engineers (SPIE) Conference Series, ed.
  {P.~J.~Quinn \& A.~Bridger}, Vol. 5493, 411

\bibitem[{{Ivanov} {et~al.}(2005){Ivanov}, {Kurtev}, \& {Borissova}}]{Ivanov05}
{Ivanov}, V.~D., {Kurtev}, R., \& {Borissova}, J. 2005, \aap, 442, 195

\bibitem[{{Joshi}(2005)}]{Joshi05}
{Joshi}, Y.~C. 2005, \mnras, 362, 1259

\bibitem[{{Kim} {et~al.}(2002){Kim}, {Demarque}, {Yi}, \& {Alexander}}]{Kim02}
{Kim}, Y.-C., {Demarque}, P., {Yi}, S.~K., \& {Alexander}, D.~R. 2002, \apjs,
  143, 499

\bibitem[{{King}(1966)}]{King66}
{King}, I.~R. 1966, \aj, 71, 64

\bibitem[{{Koposov} {et~al.}(2007){Koposov}, {de Jong}, {Belokurov}, {Rix},
  {Zucker}, {Evans}, {Gilmore}, {Irwin}, \& {Bell}}]{Koposov07}
{Koposov}, S., {de Jong}, J.~T.~A., {Belokurov}, V., {et~al.} 2007, \apj, 669,
  337

\bibitem[{{Kuchinski} \& {Frogel}(1995)}]{Kuchinski95}
{Kuchinski}, L.~E., \& {Frogel}, J.~A. 1995, \aj, 110, 2844

\bibitem[{{Kuchinski} {et~al.}(1995){Kuchinski}, {Frogel}, {Terndrup}, \&
  {Persson}}]{Kuchinski95b}
{Kuchinski}, L.~E., {Frogel}, J.~A., {Terndrup}, D.~M., \& {Persson}, S.~E.
  1995, \aj, 109, 1131

\bibitem[{{K{\"u}pper} {et~al.}(2010{\natexlab{a}}){K{\"u}pper}, {Kroupa},
  {Baumgardt}, \& {Heggie}}]{Kuepper10b}
{K{\"u}pper}, A.~H.~W., {Kroupa}, P., {Baumgardt}, H., \& {Heggie}, D.~C.
  2010{\natexlab{a}}, \mnras, 407, 2241

\bibitem[{{K{\"u}pper} {et~al.}(2010{\natexlab{b}}){K{\"u}pper}, {Kroupa},
  {Baumgardt}, \& {Heggie}}]{Kuepper10a}
{K{\"u}pper}, A.~H.~W., {Kroupa}, P., {Baumgardt}, H., \& {Heggie}, D.~C.
  2010{\natexlab{b}}, \mnras, 401, 105

\bibitem[{{Lamb} {et~al.}(1996){Lamb}, {Bulik}, \& {Coppi}}]{Lamb96}
{Lamb}, D.~Q., {Bulik}, T., \& {Coppi}, P.~S. 1996, in American Institute of
  Physics Conference Series, Vol. 366, High Velocity Neutron Stars, ed.
  {R.~E.~Rothschild \& R.~E.~Lingenfelter}, 219

\bibitem[{{Law} \& {Majewski}(2010)}]{Law10}
{Law}, D.~R., \& {Majewski}, S.~R. 2010, \apj, 718, 1128

\bibitem[{{Leitherer} {et~al.}(1999){Leitherer}, {Schaerer}, {Goldader},
  {Gonz{\'a}lez Delgado}, {Robert}, {Kune}, {de Mello}, {Devost}, \&
  {Heckman}}]{Leitherer99}
{Leitherer}, C., {Schaerer}, D., {Goldader}, J.~D., {et~al.} 1999, \apjs, 123,
  3

\bibitem[{{Longmore} {et~al.}(2011){Longmore}, {Kurtev}, {Lucas}, {Froebrich},
  {de Grijs}, {Ivanov}, {Maccarone}, {Borissova}, \& {Ker}}]{Longmore11}
{Longmore}, A.~J., {Kurtev}, R., {Lucas}, P.~W., {et~al.} 2011, \mnras, 1184

\bibitem[{{L{\'o}pez-Corredoira} {et~al.}(2002){L{\'o}pez-Corredoira},
  {Cabrera-Lavers}, {Garz{\'o}n}, \& {Hammersley}}]{Lopez02}
{L{\'o}pez-Corredoira}, M., {Cabrera-Lavers}, A., {Garz{\'o}n}, F., \&
  {Hammersley}, P.~L. 2002, \aap, 394, 883

\bibitem[{{Mallen-Ornelas} \& {Djorgovski}(1993)}]{Mallen93}
{Mallen-Ornelas}, G., \& {Djorgovski}, S. 1993, in Astronomical Society of the
  Pacific Conference Series, Vol.~50, Structure and Dynamics of Globular
  Clusters, ed. {S.~G.~Djorgovski \& G.~Meylan}, 313

\bibitem[{{Mermilliod}(1996)}]{Mermilliod96}
{Mermilliod}, J. 1996, in Astronomical Society of the Pacific Conference
  Series, Vol.~90, The Origins, Evolution, and Destinies of Binary Stars in
  Clusters, ed. {E.~F.~Milone \& J.-C.~Mermilliod}, 475

\bibitem[{{Meylan} \& {Heggie}(1997)}]{Meylan97}
{Meylan}, G., \& {Heggie}, D.~C. 1997, \aapr, 8, 1

\bibitem[{{Minniti} {et~al.}(2011){Minniti}, {Hempel}, {Toledo}, {Ivanov},
  {Alonso-Garc{\'{\i}}a}, {Saito}, {Catelan}, {Geisler}, {Jord{\'a}n},
  {Borissova}, {Zoccali}, {Kurtev}, {Carraro}, {Barbuy}, {Clari{\'a}},
  {Rejkuba}, {Emerson}, \& {Moni Bidin}}]{Minniti11}
{Minniti}, D., {Hempel}, M., {Toledo}, I., {et~al.} 2011, \aap, 527, A81

\bibitem[{{Minniti} {et~al.}(2010){Minniti}, {Lucas}, {Emerson}, {Saito},
  {Hempel}, {Pietrukowicz}, {Ahumada}, {Alonso}, {Alonso-Garcia}, {Arias},
  {Bandyopadhyay}, {Barb{\'a}}, {Barbuy}, {Bedin}, {Bica}, {Borissova},
  {Bronfman}, {Carraro}, {Catelan}, {Clari{\'a}}, {Cross}, {de Grijs},
  {D{\'e}k{\'a}ny}, {Drew}, {Fari{\~n}a}, {Feinstein}, {Fern{\'a}ndez
  Laj{\'u}s}, {Gamen}, {Geisler}, {Gieren}, {Goldman}, {Gonzalez}, {Gunthardt},
  {Gurovich}, {Hambly}, {Irwin}, {Ivanov}, {Jord{\'a}n}, {Kerins}, {Kinemuchi},
  {Kurtev}, {L{\'o}pez-Corredoira}, {Maccarone}, {Masetti}, {Merlo},
  {Messineo}, {Mirabel}, {Monaco}, {Morelli}, {Padilla}, {Palma}, {Parisi},
  {Pignata}, {Rejkuba}, {Roman-Lopes}, {Sale}, {Schreiber}, {Schr{\"o}der},
  {Smith}, {Sodr{\'e}}, {Soto}, {Tamura}, {Tappert}, {Thompson}, {Toledo},
  {Zoccali}, \& {Pietrzynski}}]{Minniti10}
{Minniti}, D., {Lucas}, P.~W., {Emerson}, J.~P., {et~al.} 2010, \na, 15, 433

\bibitem[{{Monaco} {et~al.}(2011){Monaco}, {Saviane}, {Correnti}, {Bonifacio},
  \& {Geisler}}]{Monaco11}
{Monaco}, L., {Saviane}, I., {Correnti}, M., {Bonifacio}, P., \& {Geisler}, D.
  2011, \aap, 525, 124

\bibitem[{{Moni Bidin} {et~al.}(2010){Moni Bidin}, {de La Fuente Marcos}, {de
  La Fuente Marcos}, \& {Carraro}}]{Moni09}
{Moni Bidin}, C., {de La Fuente Marcos}, R., {de La Fuente Marcos}, C., \&
  {Carraro}, G. 2010, \aap, 510, 44

\bibitem[{{Nishiyama} {et~al.}(2009){Nishiyama}, {Tamura}, {Hatano}, {Kato},
  {Tanab{\'e}}, {Sugitani}, \& {Nagata}}]{Nishiyama09}
{Nishiyama}, S., {Tamura}, M., {Hatano}, H., {et~al.} 2009, \apj, 696, 1407

\bibitem[{{Origlia} {et~al.}(2001){Origlia}, {Rich}, \& {Castro}}]{Origlia01}
{Origlia}, L., {Rich}, R.~M., \& {Castro}, S.~M. 2001, in Bulletin of the
  American Astronomical Society, Vol.~33, American Astronomical Society Meeting
  Abstracts, 1386

\bibitem[{{Origlia} {et~al.}(2005){Origlia}, {Valenti}, \& {Rich}}]{Origlia05}
{Origlia}, L., {Valenti}, E., \& {Rich}, R.~M. 2005, \mnras, 356, 1276

\bibitem[{{Ortolani} {et~al.}(2001){Ortolani}, {Barbuy}, {Bica}, {Renzini},
  {Zoccali}, {Rich}, \& {Cassisi}}]{Ortolani01}
{Ortolani}, S., {Barbuy}, B., {Bica}, E., {et~al.} 2001, \aap, 376, 878

\bibitem[{{Ortolani} {et~al.}(2009){Ortolani}, {Bonatto}, {Bica}, \&
  {Barbuy}}]{Ortolani09}
{Ortolani}, S., {Bonatto}, C., {Bica}, E., \& {Barbuy}, B. 2009, \aj, 138, 889

\bibitem[{{Pavani} \& {Bica}(2007)}]{Pavani07}
{Pavani}, D.~B., \& {Bica}, E. 2007, \aap, 468, 139

\bibitem[{{Peterson} \& {Reed}(1987)}]{Peterson87}
{Peterson}, C.~J., \& {Reed}, B.~C. 1987, \pasp, 99, 20

\bibitem[{{Rieke} \& {Lebofsky}(1985)}]{Rieke85}
{Rieke}, G.~H., \& {Lebofsky}, M.~J. 1985, \apj, 288, 618

\bibitem[{{Roeser} {et~al.}(2010){Roeser}, {Demleitner}, \&
  {Schilbach}}]{Roeser10}
{Roeser}, S., {Demleitner}, M., \& {Schilbach}, E. 2010, \aj, 139, 2440

\bibitem[{{Rosenberg} {et~al.}(1998){Rosenberg}, {Saviane}, {Piotto},
  {Aparicio}, \& {Zaggia}}]{Rosenberg98}
{Rosenberg}, A., {Saviane}, I., {Piotto}, G., {Aparicio}, A., \& {Zaggia},
  S.~R. 1998, \aj, 115, 648

\bibitem[{{Saito} {et~al.}(2010){Saito}, {Hempel}, {Alonso-Garc{\'{\i}}a},
  {Toledo}, {Borissova}, {Gonz{\'a}lez}, {Beamin}, {Minniti}, {Lucas},
  {Emerson}, {Ahumada}, {Aigrain}, {Alonso}, {Am{\^o}res}, {Angeloni}, {Arias},
  {Bandyopadhyay}, {Barb{\'a}}, {Barbuy}, {Baume}, {Bedin}, {Bica}, {Bronfman},
  {Carraro}, {Catelan}, {Clari{\'a}}, {Contreras}, {Cross}, {Davis}, {de
  Grijs}, {D{\'e}k{\'a}ny}, {Janet Drew}, {Fari{\~n}a}, {Feinstein},
  {Fern{\'a}ndez Laj{\'u}s}, {Folkes}, {Gamen}, {Geisler}, {Gieren}, {Goldman},
  {Gosling}, {Gunthardt}, {Gurovich}, {Hambly}, {Hanson}, {Hoare}, {Irwin},
  {Ivanov}, {Jord{\'a}n}, {Kerins}, {Kinemuchi}, {Kurtev}, {Longmore},
  {L{\'o}pez-Corredoira}, {Maccarone}, {Mart{\'{\i}}n}, {Masetti},
  {Mennickent}, {Merlo}, {Messineo}, {Mirabel}, {Monaco}, {Moni Bidin},
  {Morelli}, {Padilla}, {Palma}, {Parisi}, {Parker}, {Pavani}, {Pietrukowicz},
  {Pietrzynski}, {Pignata}, {Rejkuba}, {Rojas}, {Roman-Lopes}, {Ruiz}, {Sale},
  {Saviane}, {Schreiber}, {Schr{\"o}der}, {Sharma}, {Smith}, {Sodr{\'e}},
  {Soto}, {Stephens}, {Tamura}, {Tappert}, {Thompson}, {Valenti}, {Vanzi},
  {Weidmann}, \& {Zoccali}}]{Saito10}
{Saito}, R., {Hempel}, M., {Alonso-Garc{\'{\i}}a}, J., {et~al.} 2010, The
  Messenger, 141, 24

\bibitem[{{Sakamoto} \& {Hasegawa}(2006)}]{Sakamoto06}
{Sakamoto}, T., \& {Hasegawa}, T. 2006, \apjl, 653, L29

\bibitem[{{Salaris} {et~al.}(1993){Salaris}, {Chieffi}, \&
  {Straniero}}]{Salaris93}
{Salaris}, M., {Chieffi}, A., \& {Straniero}, O. 1993, \apj, 414, 580

\bibitem[{{Salaris} \& {Girardi}(2002)}]{Salaris02}
{Salaris}, M., \& {Girardi}, L. 2002, \mnras, 337, 332

\bibitem[{{Schechter} {et~al.}(1993){Schechter}, {Mateo}, \&
  {Saha}}]{Schechter93}
{Schechter}, P.~L., {Mateo}, M., \& {Saha}, A. 1993, \pasp, 105, 1342

\bibitem[{{Schlegel} {et~al.}(1998){Schlegel}, {Finkbeiner}, \&
  {Davis}}]{Schlegel98}
{Schlegel}, D.~J., {Finkbeiner}, D.~P., \& {Davis}, M. 1998, \apj, 500, 525

\bibitem[{{Skrutskie} {et~al.}(2006){Skrutskie}, {Cutri}, {Stiening},
  {Weinberg}, {Schneider}, {Carpenter}, {Beichman}, {Capps}, {Chester},
  {Elias}, {Huchra}, {Liebert}, {Lonsdale}, {Monet}, {Price}, {Seitzer},
  {Jarrett}, {Kirkpatrick}, {Gizis}, {Howard}, {Evans}, {Fowler}, {Fullmer},
  {Hurt}, {Light}, {Kopan}, {Marsh}, {McCallon}, {Tam}, {Van Dyk}, \&
  {Wheelock}}]{Skrutskie06}
{Skrutskie}, M.~F., {Cutri}, R.~M., {Stiening}, R., {et~al.} 2006, \aj, 131,
  1163

\bibitem[{{Sollima} {et~al.}(2009){Sollima}, {Bellazzini}, {Smart}, {Correnti},
  {Pancino}, {Ferraro}, \& {Romano}}]{Sollima09}
{Sollima}, A., {Bellazzini}, M., {Smart}, R.~L., {et~al.} 2009, \mnras, 396,
  2183

\bibitem[{{Stetson}(1994)}]{Stetson94}
{Stetson}, P.~B. 1994, \pasp, 106, 250

\bibitem[{{Stetson} {et~al.}(1989){Stetson}, {Hesser}, {Smith}, {Vandenberg},
  \& {Bolte}}]{Stetson89}
{Stetson}, P.~B., {Hesser}, J.~E., {Smith}, G.~H., {Vandenberg}, D.~A., \&
  {Bolte}, M. 1989, \aj, 97, 1360

\bibitem[{{Valcarce}(2011)}]{Valcarce11}
{Valcarce}, A.~A.~R. 2011, Ph.D. Thesis, Pontificia Universidad Cat\'olica de Chile

\bibitem[{{Valenti} {et~al.}(2004{\natexlab{a}}){Valenti}, {Ferraro}, \&
  {Origlia}}]{Valenti04b}
{Valenti}, E., {Ferraro}, F.~R., \& {Origlia}, L. 2004{\natexlab{a}}, \mnras,
  351, 1204

\bibitem[{{Valenti} {et~al.}(2010){Valenti}, {Ferraro}, \&
  {Origlia}}]{Valenti10}
{Valenti}, E., {Ferraro}, F.~R., \& {Origlia}, L. 2010, \mnras, 402, 1729

\bibitem[{{Valenti} {et~al.}(2004{\natexlab{b}}){Valenti}, {Ferraro}, {Perina},
  \& {Origlia}}]{Valenti04}
{Valenti}, E., {Ferraro}, F.~R., {Perina}, S., \& {Origlia}, L.
  2004{\natexlab{b}}, \aap, 419, 139

\bibitem[{{van den Bergh} {et~al.}(1991){van den Bergh}, {Morbey}, \&
  {Pazder}}]{vandenBerg91}
{van den Bergh}, S., {Morbey}, C., \& {Pazder}, J. 1991, \apj, 375, 594

\bibitem[{{VandenBerg} {et~al.}(2000){VandenBerg}, {Swenson}, {Rogers},
  {Iglesias}, \& {Alexander}}]{VandenBerg00}
{VandenBerg}, D.~A., {Swenson}, F.~J., {Rogers}, F.~J., {Iglesias}, C.~A., \&
  {Alexander}, D.~R. 2000, \apj, 532, 430

\bibitem[{{Villanova} {et~al.}(2004){Villanova}, {Carraro}, {de la Fuente
  Marcos}, \& {Stagni}}]{Villanova04}
{Villanova}, S., {Carraro}, G., {de la Fuente Marcos}, R., \& {Stagni}, R.
  2004, \aap, 428, 67

\bibitem[{{Webbink}(1985)}]{Webbink85}
{Webbink}, R.~F. 1985, in IAU Symposium, Vol. 113, Dynamics of Star Clusters,
  ed. {J.~Goodman \& P.~Hut}, 541

\bibitem[{{Wilson}(1975)}]{Wilson75}
{Wilson}, C.~P. 1975, \aj, 80, 175

\bibitem[{{Zacharias} {et~al.}(2010){Zacharias}, {Finch}, {Girard}, {Hambly},
  {Wycoff}, {Zacharias}, {Castillo}, {Corbin}, {DiVittorio}, {Dutta}, {Gaume},
  {Gauss}, {Germain}, {Hall}, {Hartkopf}, {Hsu}, {Holdenried}, {Makarov},
  {Martinez}, {Mason}, {Monet}, {Rafferty}, {Rhodes}, {Siemers}, {Smith},
  {Tilleman}, {Urban}, {Wieder}, {Winter}, \& {Young}}]{Zacharias10}
{Zacharias}, N., {Finch}, C., {Girard}, T., {et~al.} 2010, \aj, 139, 2184

\bibitem[{{Zacharias} {et~al.}(2004){Zacharias}, {Monet}, {Levine}, {Urban},
  {Gaume}, \& {Wycoff}}]{Zacharias04}
{Zacharias}, N., {Monet}, D.~G., {Levine}, S.~E., {et~al.} 2004, in Bulletin of
  the American Astronomical Society, Vol.~36, American Astronomical Society
  Meeting Abstracts, 1418

\bibitem[{{Zoccali} {et~al.}(2004){Zoccali}, {Barbuy}, {Hill}, {Ortolani},
  {Renzini}, {Bica}, {Momany}, {Pasquini}, {Minniti}, \& {Rich}}]{Zoccali04}
{Zoccali}, M., {Barbuy}, B., {Hill}, V., {et~al.} 2004, \aap, 423, 507

\end{thebibliography}

\end{document}